\documentclass[useAMS,usenatbib]{mn2e}

\usepackage{xcolor}
\usepackage{amsmath}
\usepackage{amssymb}
\usepackage{graphicx,color}
\usepackage{hyperref}

\def\aap{A\&A}

\def\mnras{MNRAS}
\def\apjl{ApJL}
\def\apjs{ApJS}
\def\apj{ApJ}

\def\nat{Nature}
\def\ext{{\rm ext}}
\def\CEsh{{\rm CE,sh}}
\def\CEin{{\rm CE,in}}
\def\CEout{{\rm CE,out}}

\newcommand{\dsfrac}[2]{\displaystyle{\frac{#1}{#2}}}

\title[Numerical models of BBD-GRBs]{Numerical models of blackbody-dominated gamma-ray bursts -- II. Emission properties}
\author[C. Cuesta-Mart\'inez, et al.]
{C. Cuesta-Mart\'inez$^{1}$\thanks{E-mail: carlos.cuesta@uv.es}, 
M. A. Aloy$^{1}$,
P. Mimica$^{1}$,
C. Th\"one$^{2}$, 
\newauthor
and
A. de Ugarte Postigo$^{2,3}$ \\
$^{1}$Departamento de Astronom\'ia y Astrof\'isica, Universidad de Valencia, E-46100 Burjassot, Valencia, Spain\\
$^{2}$Instituto de Astrof\'isica de Andaluc\'ia (IAA-CSIC), Glorieta de la Astronom\'ia s/n, E-18008 Granada, Spain\\
$^{3}$Dark Cosmology Centre, Niels Bohr Institute, Juliane Maries Vej 30, 2100 Copenhagen, Denmark}

\pagerange{\pageref{firstpage}--\pageref{lastpage}} \pubyear{}

\begin{document}

\date{\today}

\label{firstpage}

\maketitle

\begin{abstract}
Blackbody-dominated (BBD) gamma-ray bursts (GRBs) are events characterized by long durations and the presence of a significant thermal component following the prompt emission, as well as by the absence of a typical afterglow. GRB 101225A is the most prominent member of this class. A plausible progenitor system for it and for BBD-GRBs is the merger a neutron star and a helium core of an evolved, massive star. Using relativistic hydrodynamic simulations we model the propagation of ultrarelativistic jets through the environments created by such mergers. In a previous paper we showed that the thermal emission in BBD-GRBs is linked to the interaction of an ultrarelativistic jet with the ejected envelope of the secondary star of the binary. Here we focus on explaining the emission properties of BBD-GRBs computing the whole radiative signature (both thermal and non-thermal) of the jet dynamical evolution. The non-thermal emission of the forward shock of the jet is dominant during the early phases of the evolution, when that shock is moderately relativistic. Our models do not produce a classical afterglow because the quick deceleration of the jet results primarily from the mass entrainment in the beam, and not from the process of plowing mass from the external medium in front of the GRB ejecta. The contribution of the reverse shock is of the same magnitude than that of the forward shock during the first 80 min after the GRB. Later, it quickly fades because the jet/environment interaction chocks the ultrarelativistic jet beam and effectively dumps the reverse shock. In agreement with observations, we obtain rather flat light curves during the first 2 d after the GRB, and a spectral evolution consistent with the observed reddening of the system. 
\end{abstract}


\section{Introduction}
\label{sec:intro}

Gamma-ray bursts (GRBs) are flashes of $\gamma$-rays with inferred energies that link them to the most extreme relativistic and compact objects of the Universe, namely with the formation of stellar mass black holes or extremely magnetized neutron stars. The origin of the prompt $\gamma$-ray emission in GRBs is still a matter of active debate (see e.g., \citealt{Zhang_etal_2011ApJ...730..141Z}; \citealt*{Hascoet_etal_2013A&A...551A.124H}). A very direct way to obtain the observed, mostly non-thermal, spectra is assuming it is associated with the dissipation of kinetic energy in the outflow by internal shocks \citep{RM_1994ApJ}. As an alternative, the magnetic energy content of the flow can be dissipated \citep{Usov_1992Natur,Thompson_1994MNRAS,MR_1997ApJ,LB_2003astro.ph.12347L,Zhang_Yan_2011ApJ...726...90Z}. It seems very plausible that any outflow breaking out of the stellar surface will be producing a photospheric (thermal) emission (e.g., \citealt{Nakar_Sari_2012ApJ...747...88}). However, in order to produce an observed power-law tail in the high-energy part of the spectrum the photospheric emission needs to be Comptonized \citep*{Thompson_1994MNRAS,RM_2005ApJ...628..847R,GS_2007A&A...469....1G,Peer_2008ApJ...682..463P,Lazzati_etal_2009ApJ_700L_47L,Beloborodov_2010MNRAS_407_1033B}.

 Most of the detected events can be classified in any of the two standard classes proposed by \cite{Kouveliotou_etal_1993}. However, as the number of observed events increases, outliers to the basic division between short of long duration GRBs appear. Beyond their phenomenological characterization, the most interesting question that such outliers rise is whether they are typical events with atypical intrinsic and environmental physical parameters, or whether they are different types of events, perhaps associated with different channels of producing ultracompact stellar mass objects.

GRB~101225A,  also known as the `Christmas burst' (CB; \citeauthor{Thoene_etal_2011Natur_short}\,2011, hereafter T11), is a representative of the so-called blackbody-dominated GRBs (BBD-GRBs). This GRB subclass corresponds to events displaying a strong blackbody (BB) component in X-rays, which can be the dominant spectral component at optical frequencies. More recently, GRB~101225A has been linked to the new subclasses of atypically long duration GRBs, namely events with observed durations in the range $\sim 10^3 - 10^4$\,s \citep{Gendre_etal_2013ApJ...766...30G,Levan_etal_2014ApJ...781...13_short}. To account for the observational peculiarities of the CB, T11 propose that the progenitor system of this event is a merger of a neutron star with the helium core of an intermediate-mass star. Beyond the subtleties of whether the merger process results in a GRB central engine consisting of a hyperaccreting stellar mass black hole or of a powerful magnetar, the bottom line of the model is the ejection of a fraction of the hydrogen envelope of the secondary star of the binary system as a result of a quick phase of common envelope (CE) evolution of both binary members. The ejecta debris shapes the circumburst medium in such a way that the associated GRB does not display a typical afterglow and it is the reason behind the appearance of a strong thermal component in the optical bands during the first $\sim 5 - 10\,$d after the burst. We note that this scenario is not incompatible with the possibility of having additional sources of thermal emission like e.g., the X-ray emission expected in some GRBs with associated supernovae (SNe) resulting from the supernova shock breakout  \citep{Campana_etal_2006Natur_short,Soderberg_etal_2008Natur_short}. However, this X-rays contribution might be heavily modified by the absorption of the breakout flash in the ejected CE.

In \citeauthor*{Cuesta_Mimica_Aloy_2014_P1}\,(2014, hereafter Paper I),  we refine the theoretical model proposed in T11 by performing multidimensional numerical relativistic hydrodynamic simulations of jets interacting with an assumed ejecta debris. We conclude there that the origin of the thermal emission in the CB and, by extension, in other BBD-GRBs is precisely this type of jet/ejecta interaction. In the present paper we compute the non-thermal emission produced by our jet models in addition to the pure thermal emission. The synthetic light curves (LCs) and spectra obtained are compared with the observational data. This allows us to infer the true event energy, radiative efficiency and the origin of the thermal and non-thermal radiation from the CB. It also serves us for the purpose of assessing that the lack of a classical afterglow in our models is directly linked to the atypical jet/environment interaction, which produces the deceleration of the jet by mass entrainment in the beam, instead of the typical mechanism consisting of plowing mass of the external medium (EM) in front of the GRB ejecta. The main objective of this work is to explain the first $5$ d of observations. The reason we only consider this period is because, as we show in Paper I, during the first $5$ d the ultraviolet--optical--infrared (UVOIR) observations can probably be attributed chiefly to the jet--CE ejecta interaction. At later times a SN signal (T11) adds a new spectral component that we do not model in this paper. At this point it is important to stress that our goal is not to `fit' the observations with our simulations, but rather to attempt to show that, qualitatively and quantitatively, observations can be explained with reasonable and sound combinations of the physical parameters that characterize an ultrarelativistic jet and its environment.

The outline of the paper is as follows. In Section~\ref{sec:numericalmethod} we review the set-up of the hydrodynamical models in Paper I and we introduce the microphysical parameters needed to compute the non-thermal emission. In Sect.~\ref{sec:results} we present a parametric scan whose aim is to obtain the best explanation for the optical observations of GRB 101225A, while taking into account thermal and non-thermal contributions in all our models for the first 5 d of observation. In Section~\ref{sec:conclusions} we summarize the main results of our simulations and underline the strengths and weaknesses of our models. We also discuss the possible improvements in the radiative transport algorithm.

\section{Numerical method}
\label{sec:numericalmethod}

As in Paper I we use the finite volume, high-resolution shock-capturing, relativistic hydrodynamics (RHD) code \tiny{MRGENESIS }\normalsize \citep{Aloy_1999ApJS,Mimica_etal_2009ApJ}. Since we assume that the system is axisymmetric, the code is configured to run in 2D spherical coordinates. Likewise, we post-process our hydrodynamic models and compute detailed LCs and spectra with the radiative transport code \tiny{SPEV }\normalsize \citep{Mimica_etal_2004A&A...418..947,Mimica_etal_2005A&A...441..103,Mimica_etal_2009ApJ}. \tiny{SPEV }\normalsize has been extended by including the free--free bremsstrahlung process that produces the thermal emission (see Paper I). This process can account for the BB component in the observations of GRB 101225A. As remarked in Paper I, for simplicity of the thermal emission treatment, Comptonization is ignored.  In this work we also consider non-thermal emission (synchrotron radiation). In the rest of this paper we refer to the two emission processes included in the radiative code as non-thermal (synchrotron) and thermal (bremsstrahlung-BB) emission. For both emission processes we compute the observed flux in the $r$, $W2$ and $X$ bands, which correspond to frequencies of $4.68 \times 10^{14}$, $1.56 \times 10^{15}$ and $2.42 \times 10^{18}$\,Hz, respectively. Because the GRB prompt emission has been observed, and due to the probable geometry of the CE shell (with a low-density, narrow funnel along the axis), we assume that the line of sight is aligned with the rotational axis of the system and that the GRB was observed exactly head-on, i.e. the viewing angle is $\theta_{\rm obs} = 0^\circ$.

It is not possible to infer directly from observations the ultrarelativistic jet physical parameters. Therefore, as in Paper I, we first fix a reference model (RM) and the properties of the jet and of the ambient medium are varied (see Paper I for further details of the RHD numerical models). Furthermore, since the non-thermal emission also depends on a number of microphysical parameters, we consider variations of those parameters to more reliably assess the quality of our numerical results.

\subsection{Set-up of the reference model}
\label{sec:setup}

For clarity, here we briefly summarize the main properties of the hydrodynamic set-up from Paper I, and refer the reader to the schematic representation of the set-up sketched in fig.~1 of that paper. The radial grid in our simulations begins at $R_0 = 3 \times 10^{13}$ cm (the place where an ultrarelativistic jet is injected). This means that, differently from other jet simulations addressing the propagation of the jet inside of the progenitor star (e.g. \citealt{Aloy_etal_ApJL_2000__Collapsar}; \citealt*{Zhang_2003ApJ...586..356Z,Zhang_2004ApJ...608..365Z}; \citealt{Mizuta_etal_2006ApJ...651..960M}; \citealt*{Morsony_etal_2007ApJ...665..569M,Morsony_etal_2010ApJ...723..267M,Mizuta_Aloy_2009,Mizuta_etal_2011ApJ...732...26M}; \citealt{Nagakura_etal_2011ApJ...731...80N}; \citealt*{Nagakura_etal_2012ApJ...754...85N}; \citealt{Lopez_Camara_etal_2013ApJ...767...19L}), our models assume a collimated jet outside of stellar envelope, which is consistent with the results of the former models as well as with the fact that the jet is injected for a  very long time in our case. For all models with a uniform EM it ends at $R_{\rm f} = 3.27 \times 10^{15}$\,cm, and consists of 5400 uniform radial zones\footnote{For models with stratified external media or a low and uniform EM density, both $R_{\rm f}$ and the number of radial zones are larger.}. The polar grid spans the range $[0^\circ, 90^\circ]$, with a resolution of 270 uniform zones. Reflecting boundary conditions are imposed at $R_0$, at the rotational axis and at the equator. Outflow boundary conditions are set at $R_{\rm f}$.  The grid is initially filled with a cold, static, dense, uniform medium of density $\rho_{\ext} = 8 \times 10^{-14}$\,g\,cm$^{-3}$. Starting at a distance $R_{\CEin} = 4.5 \times 10^{13}$\,cm and ending at $R_{\CEout} = 1.05 \times 10^{14}$\,cm we place a high-density uniform CE shell at rest. The shell possesses a low-density funnel around the symmetry axis with an opening angle of $\theta_{\rm f,in} = 1^\circ$ at $r = R_{\CEin}$. The opening angle either grows exponentially with radius (to reproduce a toroidal-like shape), or linearly (see fig.~1 of Paper I) and reaches $\theta_{\rm f,out} = 30^\circ$ at $r = R_{\CEout}$. For the CE-shell density we take $\rho_\CEsh = 1.2 \times 10^{-10}$\,g\,cm$^{-3}$, so that $\rho_\CEsh = 1500 \rho_{\ext}$ and $p_\CEsh/\rho_\CEsh \approx 6.7 \times 10^{-9}c^2$ ($c$ being the speed of light in vacuum). This density corresponds to an ejecta mass $\sim 0.26\,M_\odot$. In models with a uniform density in the EM, the pressure in the circumburst medium and in the CE shell is uniform ($p_\ext = p_\CEsh$) and we set it low enough that the ambient medium is a cold (non-relativistic) plasma ($p_\ext/\rho_{\ext} = 10^{-5}c^2$). This low ambient pressure only exerts a tiny influence on the jet dynamics during the initial 5 d of evolution.  For the RM we have chosen a jet opening angle of $\theta_{\rm j} = 17^\circ$, ensuring that the jet beam is wider than the funnel when it hits the innermost radial boundary of the CE shell. The jet has an initial Lorentz factor $\Gamma_{\rm i} = 80$, and its specific enthalpy is set to $h_{\rm i} = 5$, so that, if allowed to propagate without obstacles, it can accelerate to an asymptotic Lorentz factor $\Gamma_\infty \approx \Gamma_{\rm i} h_{\rm i} = 400$. We choose an isotropic equivalent total energy of the jet $E_{\rm iso} = 4 \times 10^{53}$\,erg for our RM. The jet is injected in the computational domain in two consecutive phases\footnote{In \cite{Aloy_etal_2013ASPC..474...33} we assumed a unique constant injection interval with duration equal to $t_{\rm inj}$.}: (1) constant up to $T_{\rm1}\simeq1100\,$s and (2) variable with a time dependence $\propto t^{-5/3}$ (both in $\rho$ and in $p$, so that the ratio $p/\rho$ is constant) up to $T_{\rm2}\simeq 3800\,$s. For reasons of numerically stability, at times $T>T_{\rm2}$ the jet is progressively switched off (by reducing both the injected rest mass and pressure as $\propto t^{-4}$) rather than turned off abruptly.

\subsubsection{Non-thermal microphysical parameters}\label{sec:microparams}

The non-thermal microphysical parameters for the calculation of the synchrotron emission have been chosen so that the sum of the thermal and the non-thermal emission is consistent with the observations. Behind the forward shock a fraction of the fluid internal energy is carried by relativistic electrons ($\epsilon_{\rm e} = 10^{-3}$) and by stochastic magnetic fields ($\epsilon_B = 10^{-6}$; see Appendix~\ref{sec:app.nonthermal}). For a large number of GRB afterglows, observations suggest a value $\epsilon_B\lesssim 10^{-4}$ \citep{Kumar_etal_2012MNRAS.427L..40}. We choose an even smaller value for $\epsilon_B$ for several reasons. On the one hand, $\epsilon_B$ can grow from the lower values in the interstellar medium (ISM) to $\epsilon_B\sim 0.1$ due to the amplification of the magnetic field by the two-stream instability \citep{ML_1999ApJ...526..697}. However, this field quickly decays away from the shock in a few plasma scales (e.g., \citealt{SS_2011ApJ...726...75}), and it is difficult to estimate its value far from the shock front. \cite{SS_2011ApJ...726...75} find in their PIC models that the far field strength should be of the order of the shocked ISM field strength assuming flux freezing. Provided that our typical grid size is $\sim 10^{11}$--$10^{12}\,$cm, over the first post-shock numerical cell the stochastic magnetic field may have decayed to the compressed ISM field values estimated by \cite{SS_2011ApJ...726...75}. Since our ISM is assumed to be unmagnetized, a value $\epsilon_B \sim 10^{-5}$--$10^{-6}$ is a plausible `average' of $\epsilon_B$ in the post-shock state. On the other hand, to be consistent with the assumption that the magnetic field be dynamically negligible (i.e. to be consistent with our fluid dynamical approach), we require that $\epsilon_B <10^{-2}$ \citep*[e.g.][]{Mimica_etal_2007A&A...466...93}. And, finally, large values of $\epsilon_B$ yield very high stochastic magnetic fields ($B_{\rm st} \ga 10^4$\,G) that would cause superfast synchrotron cooling of the electrons. Such short cooling time-scales cannot be resolved using our simulations, because we would need to resolve in our explicit hydrodynamic models time-scales which are shorter than the synchrotron cooling time of the electrons contributing to radiation in the observing bands, which is $t_{\rm cool}\propto \epsilon_B^{-3/4}$. For the numerical resolutions employed in this paper, the typical time step is $\simeq 10\,$s. With the value $\epsilon_B=10^{-6}$, the cooling time of electrons contributing to the radiation in the X-ray band is $\simeq 500\,$s. Increasing $\epsilon_B$ by a factor of 100 lowers that cooling time to $\simeq 20\,$s. This time-scale could be resolved by the hydrodynamical time resolution of our simulation, but only if we would post-process and record \emph{every} time step. Unfortunately, at present this is not feasible since we would need to record more than $4\times 10^4$ files per 2D model. Of course, if we limit ourselves to the evaluation of the non-thermal flux density at optical frequencies, we could safely consider values $\epsilon_B\lesssim 10^{-3}$. Beyond these technical limitations, $\epsilon_B$ could be different for different shocks, but we assume, for simplicity, that it is the same in all cases.
%
%

The number of relativistic electrons accelerated by the shock is a fraction, $\zeta_{\rm e} = 0.1$, of the total number of electrons in the pre-shocked state. The power-law index of the electron energy distribution is $p = 2.3$, and the acceleration efficiency parameter, $a_{\rm acc} = 1$ (see Appendix~\ref{sec:app.nonthermal}). We can define an effective fraction of internal energy for the electrons, $\epsilon'_{\rm e} = \epsilon_{\rm e} / \zeta_{\rm e} $. With such definition,  $\epsilon'_{\rm e} \approx 0.01$, i.e. closer to the `typical' values of $\epsilon_{\rm e}$ commonly used in other studies in which the $\zeta_{\rm e}$ parameter is absent. A similar parameterization can be found in \cite*{BDD_2009A&A...498..677}. 

\begin{table*}
\centering
\caption{Summary of the most important thermal and the non-thermal properties that describe the models. For each model, we indicate in bold the parameter that is different from the RM. The equivalent isotropic energy is expressed in units of $10^{53}$\,erg. The column `Geometry' refers to the geometrical shape of the CE shell, and models where the shell has a toroidal shape are annotated with `T', and those in which the funnel is linear with `L'.   In the column `Ext. medium', models with a uniform EM are annotated with `U'  (`U1' denotes a density $\rho_{\rm ext} = 8 \times 10^{-14}$\,g\,cm$^{-3}$ and `U2' a density  $\rho_{\rm ext} = 8 \times 10^{-15}$\,g\,cm$^{-3}$), and models with a stratified medium with `S'.}
\label{tab:params}
\begin{tabular}{|l|c|c|c|c|c|c|c|c|c|c|c|c|c|c|c|c|c|c|}
\hline Model & $\theta_{\rm j}$ & $E_{{\rm iso},53}$ & $\rho_\CEsh / \rho_{\ext}$ & Geometry & $\theta_{\rm f,out}$ & Ext. medium & $p$ & $\epsilon_{\rm e}/10^{-3}$ & $\epsilon_B/10^{-6}$ & $\zeta_{\rm e}$ & $a_{\rm acc}$ \\
\hline  RM & $17^\circ$ & 4 & 1500 &  T & $30^\circ$ & U1 & 2.3 & 1 & 1 & $0.1$ & 1\\
	T14 & $\boldsymbol{14^\circ}$ & 4 & 1500 &  T & $30^\circ$  & U1 & 2.3 & 1 & 1 & $0.1$ & 1\\
	T20 & $\boldsymbol{20^\circ}$ & 4 & 1500 &  T & $30^\circ$ & U1 & 2.3 & 1 & 1 & $0.1$ & 1\\
	E53 & $17^\circ$ & \textbf{2} & 1500 & T & $30^\circ$ & U1 & 2.3 & 1 & 1 & $0.1$ & 1\\
	D2 & $17^\circ$ & 4 & \textbf{817} & T & $30^\circ$ & U1 & 2.3 & 1 & 1 & $0.1$ & 1\\
	GS & $17^\circ$ & 4 & \textbf{4304} & \textbf{T$^{(a)}$} & $30^\circ$ &  U1 & 2.3 & 1 & 1 & $0.1$ & 1\\
	G2 & $17^\circ$ & 4 & 1500 &\textbf{L} & $30^\circ$ & U1 & 2.3 & 1 & 1 & $0.1$ & 1\\
	G3 & $17^\circ$ & 4 & 1500 & T & $\boldsymbol{15^\circ}$ & U1 & 2.3 & 1 & 1 & $0.1$ & 1\\
	M2 & $17^\circ$ & 4 & \textbf{15 000} & T & $30^\circ$ & \textbf{U2$^{(b)}$} & 2.3 & 1 & 1 & $0.1$ & 1\\
	S1 & $17^\circ$ & 4 & 1500 &  T & $30^\circ$ & \textbf{S$^{(c)}$} & 2.3 & 1 & 1 & $0.1$ & 1\\
	S2 & $17^\circ$ & 4 & 1500 &  T & $30^\circ$ & \textbf{S$^{(d)}$} & 2.3 & 1 & 1 & $0.1$ & 1\\
	P4 & $17^\circ$ & 4 & 1500 &  T & $30^\circ$ & U1 &  \textbf{2.4} & 1 & 1 & $0.1$ & 1\\
	P5 & $17^\circ$ & 4 & 1500 &  T & $30^\circ$ & U1 &  \textbf{2.5} & 1 & 1 & $0.1$ & 1\\
	P6 & $17^\circ$ & 4 & 1500 &  T & $30^\circ$ & U1 &  \textbf{2.6} & 1 & 1 & $0.1$ & 1\\
	EE2 & $17^\circ$ & 4 & 1500 & T & $30^\circ$ & U1 &  2.3 & $\boldsymbol{10}$ & 1 & $0.1$ & 1 \\
	EB5 & $17^\circ$ & 4 & 1500 & T & $30^\circ$ & U1 &  2.3 & 1 & $\boldsymbol{10}$ & $0.1$ & 1 &\\
	ZE2 & $17^\circ$ & 4 & 1500 & T & $30^\circ$ & U1 &  2.3 & 1 & 1 & $\boldsymbol{0.01}$ & 1\\
	A6 & $17^\circ$ & 4 & 1500 &  T & $30^\circ$ & U1 &  2.3 & 1 & 1 & $0.1$ & $\boldsymbol{10^6}$\\\hline
\end{tabular}
\\$^{\rm (a)}$ In model GS, the CE-shell rest-mass density and pressure are not uniform but decay with $r^{-2}$.
\\$^{\rm (b)}$ In model M2, the pressure in the EM and CE shell is $p_{\rm ext}/\rho_{\rm ext} = 10^{-4}c^2$.
\\$^{\rm (c)}$ In model S1, the EM has a rest-mass density and pressure that decay with $r^{-1}$.
\\$^{\rm (d)}$ In model S2, the EM has a rest-mass density and pressure that decay with $r^{-2}$.
\end{table*}

\section{Results}
\label{sec:results}

In this section we examine the synthetic LCs and spectra obtained with the \tiny{SPEV }\normalsize code. In contrast to Paper I, we include here both the thermal and the non-thermal contributions to the whole spectral signature of our models. LCs are computed in three different bands, $r$, $W2$ and X-rays. Other intermediate frequencies among the previous ones have also been computed in order to obtain synthetic spectra.  We first discuss the results for the RM (Sect.~\ref{sec:referencemodel}) and then consider variations of the parameters in Sect.~\ref{sec:parametricscan}. A summary of the most relevant parameters of the models is given in Table~\ref{tab:params}.

\subsection{Origin of the non-thermal emission}
\label{sec:origin-nonthermal}
\begin{figure}
\centering
\includegraphics[width=8.4cm]{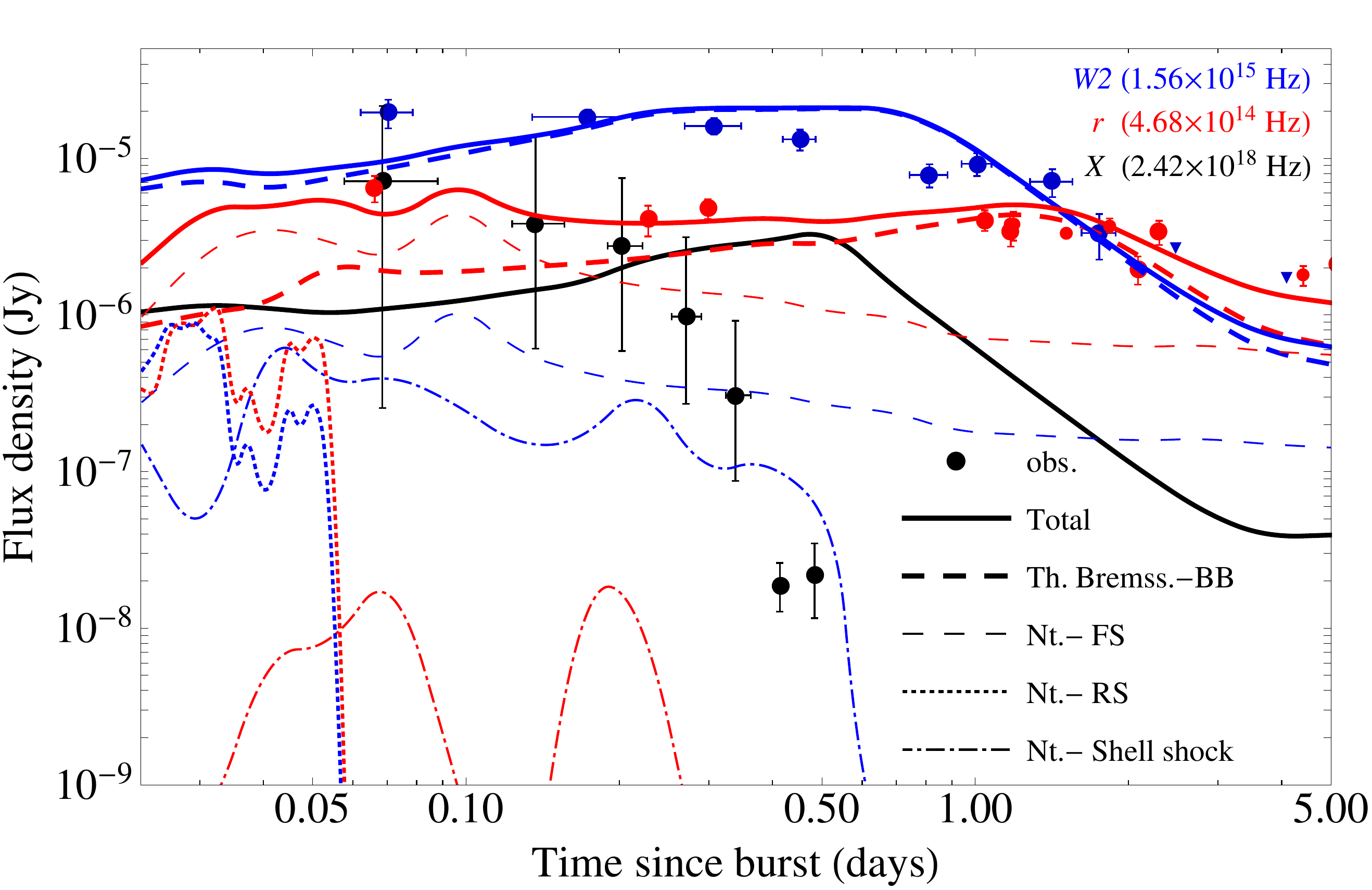}
\caption{Synthetic LCs during the first 5 d for RM. We show the total emission (solid lines) and the individual contribution of the thermal (thick dashed lines) and various non-thermal radiation sources. For the latter case, we distinguish the contributions arising from the shocks resulting from the jet/CE-shell interaction (thin dot--dashed lines), the forward shock (thin dashed lines) and the reverse shock (short-dashed lines). The observational data have been taken from T11 and references therein (large solid circles), and from \citeauthor{Levan_etal_2014ApJ...781...13_short} (2014; small solid circles). Upper observational limits are represented as triangles. Red, blue and black colours are used to display data in the $r$, $W2$ and X-ray bands, respectively. The non-thermal X-ray band flux arising from shocks resulting from the jet/CE-shell interaction is very absorbed and cannot be shown on the scale we are representing in the figure.}
\label{fig:NTcontributions}
\end{figure}
\begin{figure}
\centering
\includegraphics[width=8.4cm]{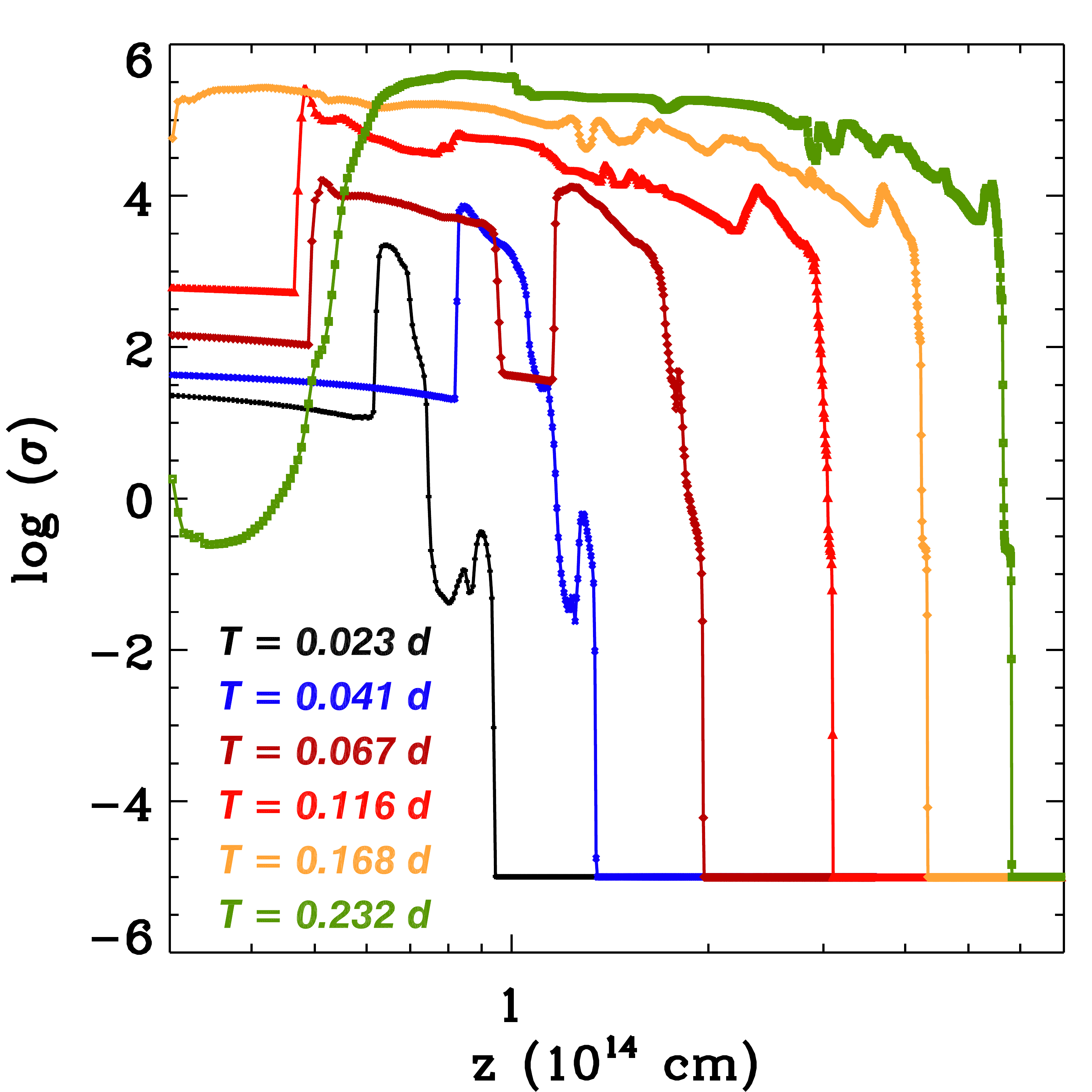}
\caption{Profiles of the logarithm of the entropy density along the symmetry axis at different evolutionary lab-frame times (see legends). Since $\sigma \propto p / \rho^\gamma$ ($\gamma$ being the adiabatic index, which in our case runs between $4/3$ and $5/3$), and since we keep $p/\rho=$constant at the injection nozzle, the entropy density will grow in time in the switch off phase, namely, after $T=0.023\,$d as $\sigma(t) \propto t^{5(\gamma-1)/3}$. }
\label{fig:entropy}
\end{figure}

In our models the non-thermal emission is produced by particles accelerated at (strong) shocks. As in any weakly magnetized or unmagnetized jet model, there are at least two main shocks produced by the jet/medium interaction. These are the forward (FS) and reverse shocks (RS) or, using a terminology more common in jet hydrodynamics, the bow shock and the Mach disc. Since the jet is not pressure matched, the beam of the jet may also develop a number of bi-conical shocks where the flow is pinched. In the specific scenario we are considering, the jet blasts the CE shell and drives two main shocks, qualitatively similar to the FS and RS, but moving towards the jet axis (thus, shocking the jet beam) and away from it (compressing the CE shell). Though, in principle all of these shocks may accelerate non-thermal electrons and, thereby contribute to the synchrotron emission of our models, we will justify in the following that the dominant synchrotron flux arises the observer from the FS.

Bi-conical (recollimation) shocks are much weaker than the FS or the RS, and the efficiency of shock acceleration in such oblique shocks is smaller than that of shocks perpendicular to the fluid motion. Thus, we can safely ignore them when accounting for the non-thermal emission.

Regarding the shocks driven by the jet/CE-shell interaction, they move in a high-density medium at a relatively small angle with respect to the jet axis. This part of the system is opaque to the synchrotron radiation during most of the evolution considered here. We note that the shell density is so high that the optical depth of the fluid through which the shocks are propagating is $\tau_\nu\gtrsim 10$ for all frequencies under consideration. To completely assess the contribution of shocks resulting from the jet/CE-shell interaction, we have included them in the computation of the non-thermal emission of the RM. As can be seen in Fig.~\ref{fig:NTcontributions} (dot--dashed lines), their contribution is only significant (but still subdominant) until $t_{\rm obs}\lesssim 0.6\,$d, and only in the $W2$ band. By this time we are no longer able to detect these shocks, since they are mixed with a much distorted and partly ablated CE shell. In the $r$ band, and specially in the X-ray band (which cannot be seen on the scale of the figure), their contribution to the total flux density is absolutely negligible.

During the time interval when the injected jet power is sufficiently high (e.g. until $0.058$\,d) we see a distinct entropy density jump of more than two orders of magnitude on both the FS and the RS  (Fig.~\ref{fig:entropy}). These shocks are located at $z_{\rm FS}\simeq 9.2\times 10^{13}\,$cm and $z_{\rm RS}\simeq 6.2\times 10^{13}\,$cm at $T \simeq 0.023\,$d, and at $z_{\rm FS}\simeq 1.3\times 10^{14}\,$cm and $z_{\rm RS}\simeq 8\times 10^{13}\,$cm at $T \simeq 0.041\,$d (Fig.~\ref{fig:entropy}, black and blue lines). However, not long after the jet head leaves the outer edge of the CE shell, two oppositely moving shocks driven by the jet/CE-shell interaction, travel towards the jet axis and eventually met. The brown line in Fig.~\ref{fig:entropy}, corresponds to a time ($T \simeq 0.067\,$d) at which these two shocks have reached the jet beam. We note the rise of the entropy density in the unshocked tail of the jet (i.e. the part of the jet trailing RS that is free expanding). As a result of the jet/CE-shell interaction the jet beam is choked, strongly baryon loaded and the RS is smeared so much that its strength is greatly decreased. In practical terms there is effectively no entropy jump across the locus of the RS after $T =0.116\,$d (when the RS is situated at $z \simeq 2.3\times 10^{14}\,$cm; Fig.~\ref{fig:entropy}, red line). Instead, we find a steep positive entropy density gradient.  Progressively, the high entropy of the matter coming from the shocked CE shell fills the unshocked jet tail, eventually reaching the inner grid boundary (Fig.~\ref{fig:entropy}; $T =0.168\,$d). In short, the jet/CE-shell interaction effectively blurs the RS after a lab-frame time $T \simeq 0.07\,$d. Hence, any radiative signature of the RS may only influence our emission results at early times. Indeed, we have computed the non-thermal flux associated with the RS until a bit after the time in which we can still identify in our models a RS, and we notice that, in the $W2$ band, the RS signature is one order of magnitude below that of the FS (Fig.~\ref{fig:NTcontributions}; blue short-dashed line). The situation in the $r$ band is quantitatively different. In this band, the RS non-thermal flux emission is comparable to that of the FS. The relative contribution of the RS is more relevant at smaller frequencies. Unfortunately, the observational data in the $z$ band is scarce, and available only after $\simeq 0.3\,$d after the burst (T11). Hence, we cannot constrain our models with observational data in a band where the RS would be dominant over the FS.

\begin{figure*}
\centering
\centering
\includegraphics[width=18cm]{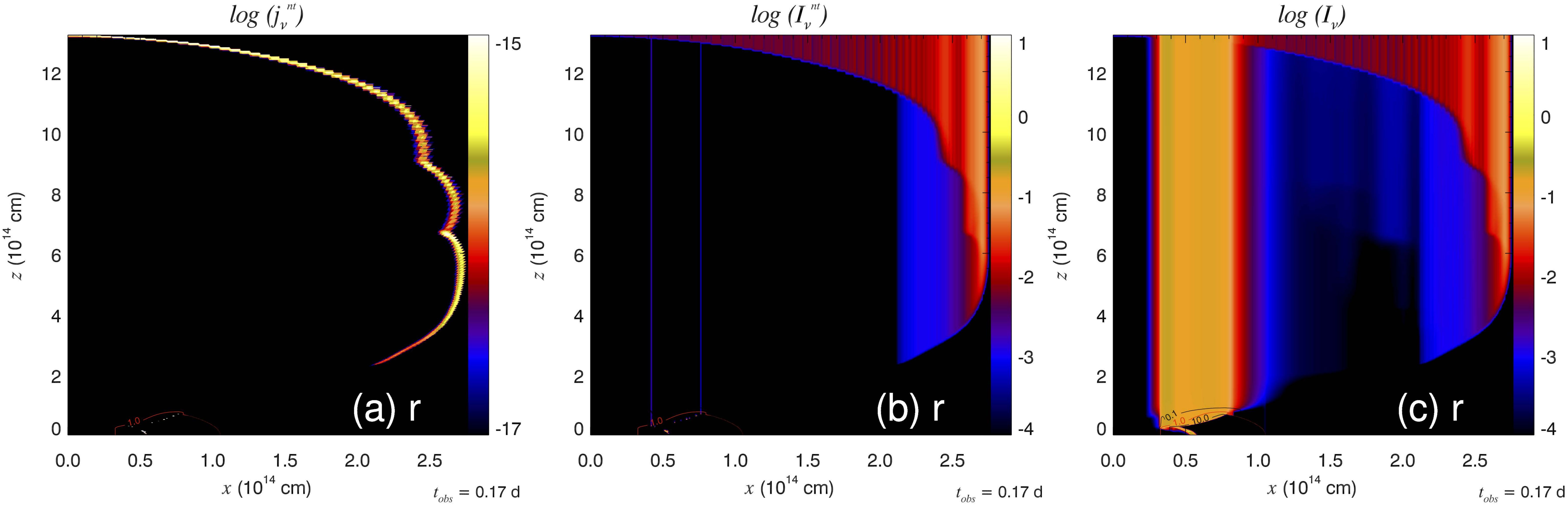}
\includegraphics[width=18cm]{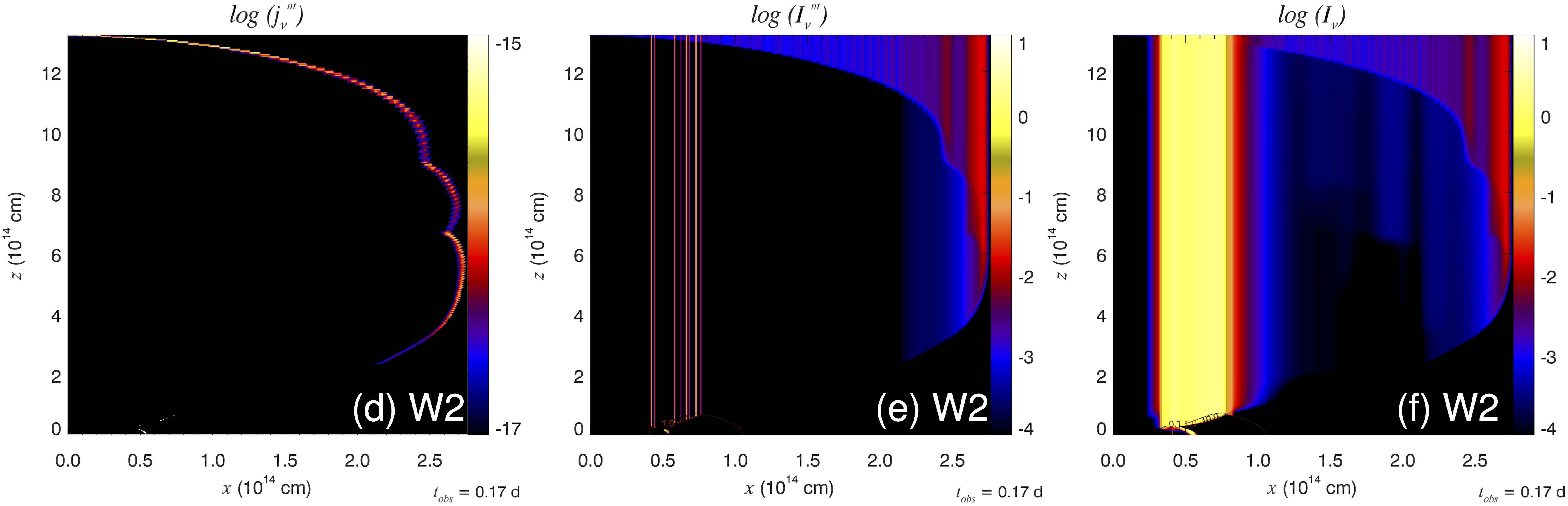}
\includegraphics[width=18cm]{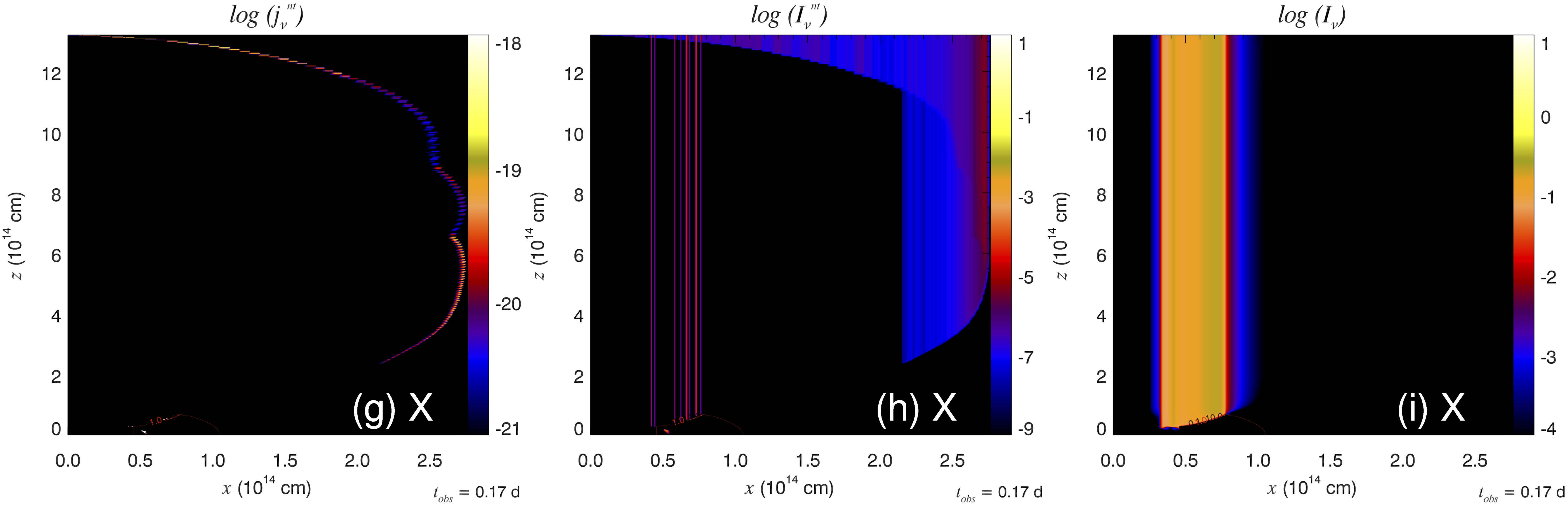}
\caption{Synchrotron emissivity, $j_\nu$ (left-hand column), and evolution of the specific intensity, $I_\nu$, along the line of sight considering only non-thermal processes (central column) or both (thermal and non-thermal) processes simultaneously (right-hand column). The observer is located in the vertical direction (towards the top of the page) at a viewing angle $\theta_{\rm obs} = 0^\circ$. The emission is computed in the $r$ (upper row), $W2$ (middle row) and X-rays (bottom row) bands at an observational time $t_{\rm obs} = 0.17$\,d. Note that, while the  colour scales are the same for the total specific intensity (right-hand column), they are different for the lower left and central panels, since the non-thermal X-ray emission is much fainter than in the other two bands displayed in the figure. The units of $j_\nu$ and $I_\nu$ are given in the CGS system (see Appendix~\ref{sec:spev} for details).}
\label{fig:RM-W2-TH-NT}
\end{figure*}

Turning now to Fig.~\ref{fig:RM-W2-TH-NT} we clearly see that the production sites of the observed non-thermal and thermal radiation observed are distinct. The main contribution to the total thermal intensity is due to radiation originating from the jet/CE-shell interaction region, located at a distance from the symmetry axis of $\simeq 3\times 10^{13}\,$cm and extending to $\simeq 8\times 10^{13}\,$cm. In the specific intensity maps in Fig.~\ref{fig:RM-W2-TH-NT} (panels a, d and g) such a dominant contribution is clearly seen as a light (yellow or orange) fringe emerging from the jet/CE-shell interaction region. This is the case in all the three bands considered here. On the other hand, the non-thermal emission comes from the bow (forward) shock of the jet (Fig.~\ref{fig:RM-W2-TH-NT}, panels b, e and h). We also can see the contribution of the shocks driven in the CE shell as a set of intense vertical fringes in the range $4\times 10^{13}<x<8\times 10^{13}\,$cm. In spite of its large intensity density, the contribution of these CE-shell shocks to the total flux is smaller than that of the FS at almost any time.

Since the synchrotron self-absorption is very low at these frequencies, the forward shocked region is optically thin to the non-thermal radiation and, hence, the observed non-thermal specific intensity follows closely the locus of the forward shock in the observer's frame (Fig.~\ref{fig:RM-W2-TH-NT}, panels b, e and h). We point out that in the maps displayed in Fig.~\ref{fig:RM-W2-TH-NT}, we have not considered the contribution of the RS which is negligible at $t_{\rm obs}=0.17\,$d. As a result of the jet/CE-shell interaction, the forward shock is not perfectly spherical. Instead, even at 0.17\,d, we can still see a number of kinks in the shocked emission surface between $2.4\times 10^{14}\,$ and  $2.7\times 10^{14}\,$cm. Furthermore, we also note that since the non-thermal emission results from the outer (geometrically thin) skin of the cavity, the non-thermal emission region appears limb brightened for the observer. We also observe that the thermal and non-thermal contributions to the total flux are much more evenly distributed in the $r$ band than in the $W2$ band.

\begin{figure}
\centering
\includegraphics[width=8.4cm]{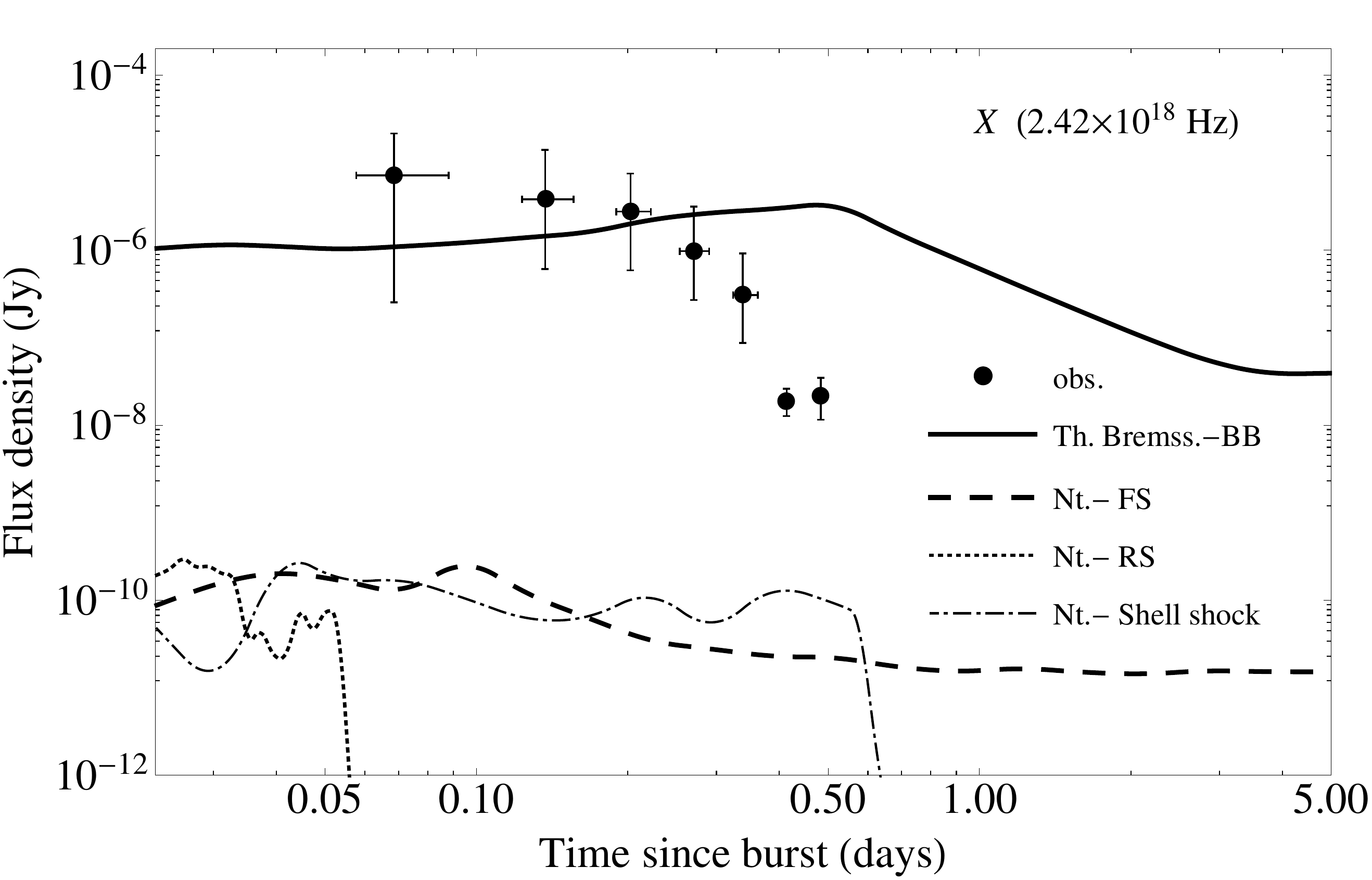}
\caption{Synthetic LC in X-ray band during the first 5 d for RM. We show the thermal (solid lines) and the non-thermal (dashed lines) radiation. For the representation of the X-ray data we have clustered the data of the X-Ray Telescope (XRT) observing cycle into a single point, with error bars showing the data dispersion.}
\label{fig:RM-X}
\end{figure}

\begin{figure}
\centering
\centering
\includegraphics[width=8.4cm]{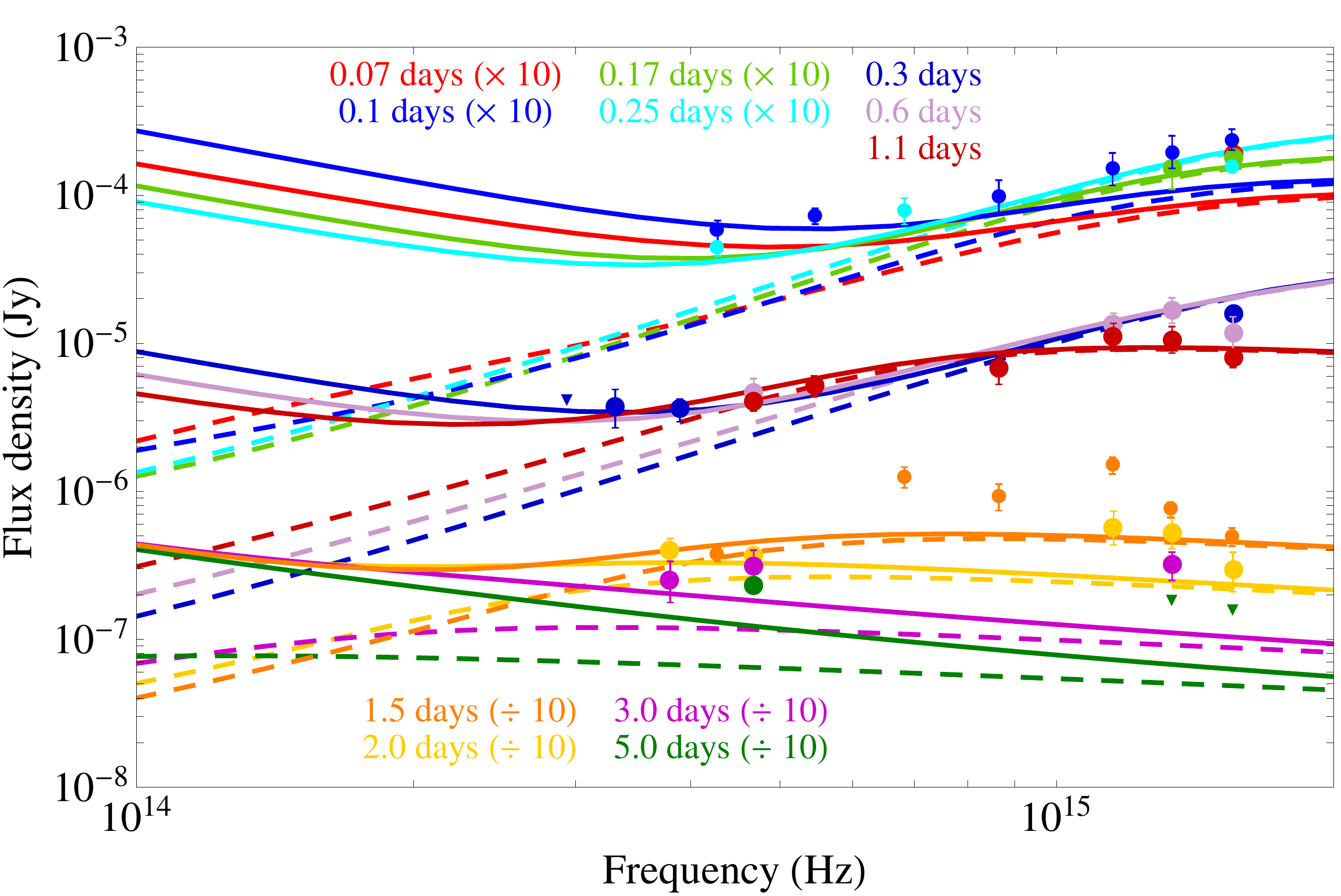}
\caption{Fitted spectrum at different observing times (colours denote observations at different times, see legend; detections at $0.1$, $0.25$ and $1.5$ d are taken from \citealt{Levan_etal_2014ApJ...781...13_short}; rest of data taken from T11) for the RM. The total contribution (non-thermal plus thermal) is represented as solid lines and the thermal contribution as dashed lines. Note that for visualization convenience some of the data have been multiplied or divided by a factor of 10 (see the plot legends).}
\label{fig:spectrum}
\end{figure}
%

\subsection{Reference model}
\label{sec:referencemodel}

As we have already shown in the previous section, the synthetic  LCs computed in the $W2$ and $r$ bands are very close to the observational points (Fig.~\ref{fig:NTcontributions}). As anticipated in Sect.~\ref{sec:intro}, our models are not a `perfect fit' of the observations, even for the RM, though they qualitatively and quantitatively agree with them. At early times, the $W2$ band is more energetic than the $r$ band, in agreement with observational data.  The contribution of the non-thermal emission to the total LC is significant at early times ($t_{\rm obs}\la 0.5$\,d; Fig.~\ref{fig:NTcontributions}), and is more relevant at lower frequencies. It is dominant, by a factor of $< 2$, in the $r$ band until $\approx 0.15\,$d. Looking at the distribution of the specific intensity in the $r$ band (Fig.~\ref{fig:RM-W2-TH-NT}c), this fact is not due to a much larger specific (synchrotron) intensity at the FS, but to the fact that the FS surface is larger than the thermally emitting region.  In the $W2$ band, the thermal emission is larger at any time than the non-thermal one and, indeed, the thermal flux density is more than one order of magnitude larger than the non-thermal one after $t_{\rm obs}=0.17\,$d (Fig.~\ref{fig:NTcontributions}). The bremsstrahlung-BB contribution represents $93$ and $63$ per cent of the total received flux in the $W2$ and in the $r$ band, respectively.

In the X-ray band, the non-thermal flux is much smaller than the thermal one at $t_{\rm obs}=0.17\,$d (Fig.~\ref{fig:RM-X}), which is due to the fact that the X-ray emission of the FS is extremely weak (note the difference in the scales on the left-hand panels of Fig.~\ref{fig:RM-W2-TH-NT}). This happens because the FS is only moderately relativistic at this time of the jet evolution. In many regards, the emission of the FS resembles that of the very late afterglow in a standard GRB.

At late times, our models suggest that the non-thermal emission might eventually be relevant. In the $r$ band, the non-thermal flux can be comparable in magnitude to the thermal one after $t_{\rm obs}\gtrsim 4\,$d.  The same qualitative feature may happen much later in the $W2$ band, since the difference between the thermal and non-thermal contributions to the flux density decreases progressively until there is only a factor of $\sim 4$ at $t_{\rm obs}=5\,$d.

The agreement between our model and the observations is not limited to the LC, but also extends to the spectral data (Fig.~\ref{fig:spectrum}) for the first 5 d after the burst. The agreement between observed spectra and the synthetic ones computed for our RM is much better at early and intermediate times than after $\sim 3$\,d. The thermal contribution is clearly dominant for frequencies $\ga 6 \times 10^{14}$\,Hz. In the X-ray band the non-thermal contribution is practically negligible (Fig.~\ref{fig:RM-X}).
\begin{figure}
\centering
\centering
\includegraphics[width=8.4cm]{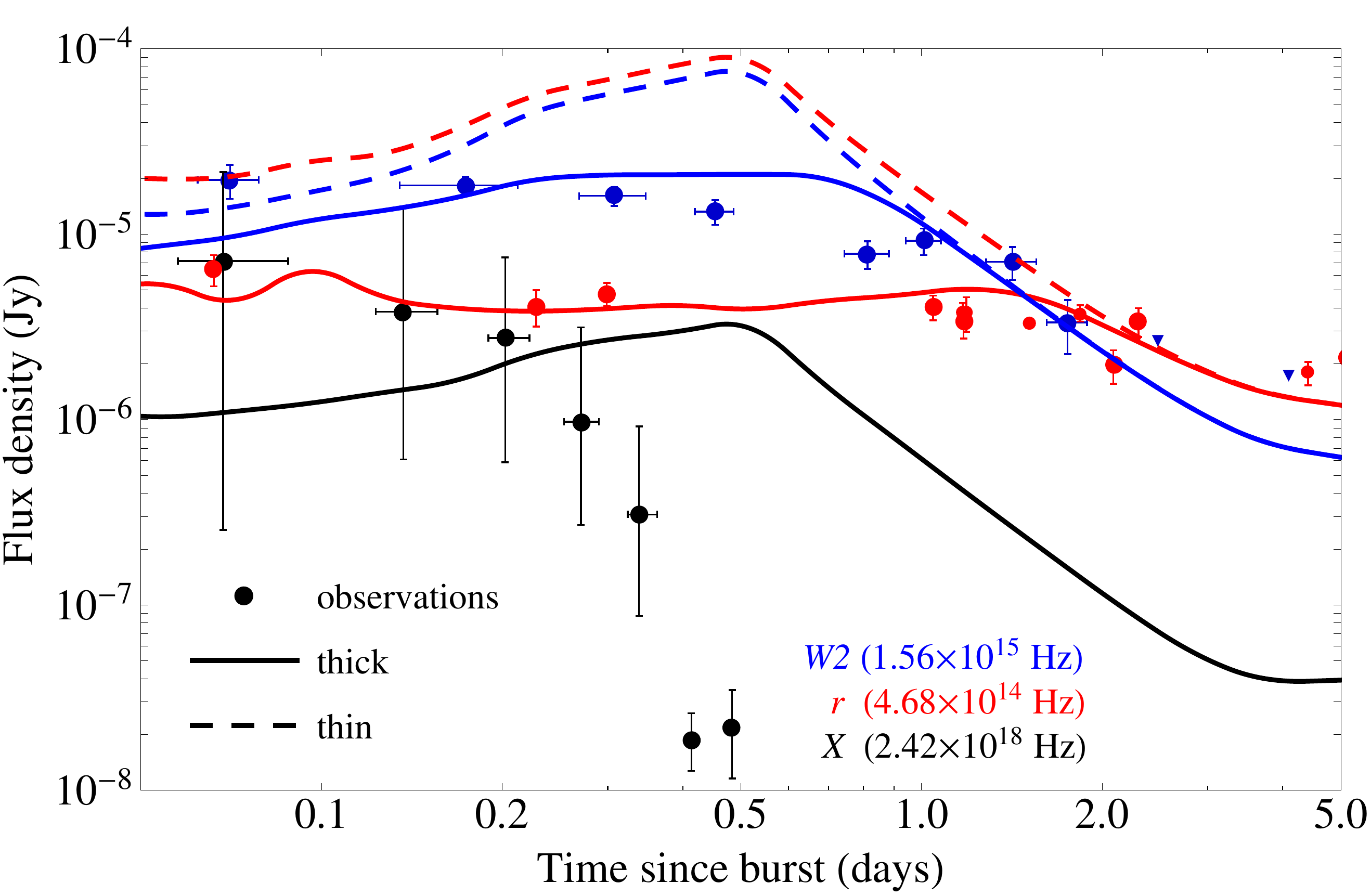}
\caption{Optically thick (solid lines) and thin (dashed lines) total light curves in the $W2$, $r$ and $X$-ray band are plotted to illustrate the transition from optically thin to optically thick emission. It is useful to compare this figure with Fig.~11 in Paper I, where only the thermal contribution is considered. Observational data is marked with symbols (see Fig.~\ref{fig:NTcontributions}).}
\label{fig:referencethickthin}
\end{figure}
Even when we include the non-thermal contributions to the emitted radiation, the system is optically thick until $\sim 1$\,d after the burst in the $W2$ band, and until $\sim 2$\,d in the $r$ band (Fig.~\ref{fig:referencethickthin}). As we also saw in Paper I, the system is optically thin in the X-ray band at all times. After $\sim 1.5$--$2$ d, when the system becomes optically thin at all optical frequencies, it experiences a spectral inversion (we begin to observe more flux in the $r$ band than in the $W2$ band). This feature is in remarkable agreement with observations and, as we have shown in Paper I, it is related to the time by which the CE shell is effectively ablated by the ultrarelativistic jet.

\subsection{Parametric scan}
\label{sec:parametricscan}

We have found that the parameters of the RM are an appropriate fit to the observations, but it turns out that this combination of the parameters is not the only one able to satisfactorily explain them. As we have pointed out in Section \ref{sec:referencemodel}, at lower frequencies the non-thermal contribution is more relevant, so any modification in any of the non-thermal microphysical parameters will be reflected to a greater degree in the $r$ band than in $W2$. We have shown in Sect.~\ref{sec:referencemodel} for the RM that the X-ray non-thermal contribution can be neglected since it is $\sim 4$ orders of magnitude smaller than the thermal contribution. This conclusion can be extended to the rest of the models in this paper since, as pointed out in the section 5.2 of Paper I, the thermal X-ray emission in all our models displays a (thermal) flux deficit at early times that cannot be compensated by the non-thermal X-ray flux. Furthermore, the duration of the X-ray emission is also too long to accommodate the observational data (see fig.~20 of Paper I). We concluded that to reproduce the observations we shall tune the properties of the CE shell (e.g. the density gradient both in the radial and angular directions), something which would require also much more numerical resolution and prohibitively long computing times to perform a broad parametric scan of sufficiently representative models. Therefore, we will not focus the parametric scan in this band in the rest of the paper.

Table~\ref{tab:params} lists the parameters of all the models we are considering here.

\subsubsection{Non-thermal microphysical parameters}

\begin{figure}
\centering
\includegraphics[width=8.4cm]{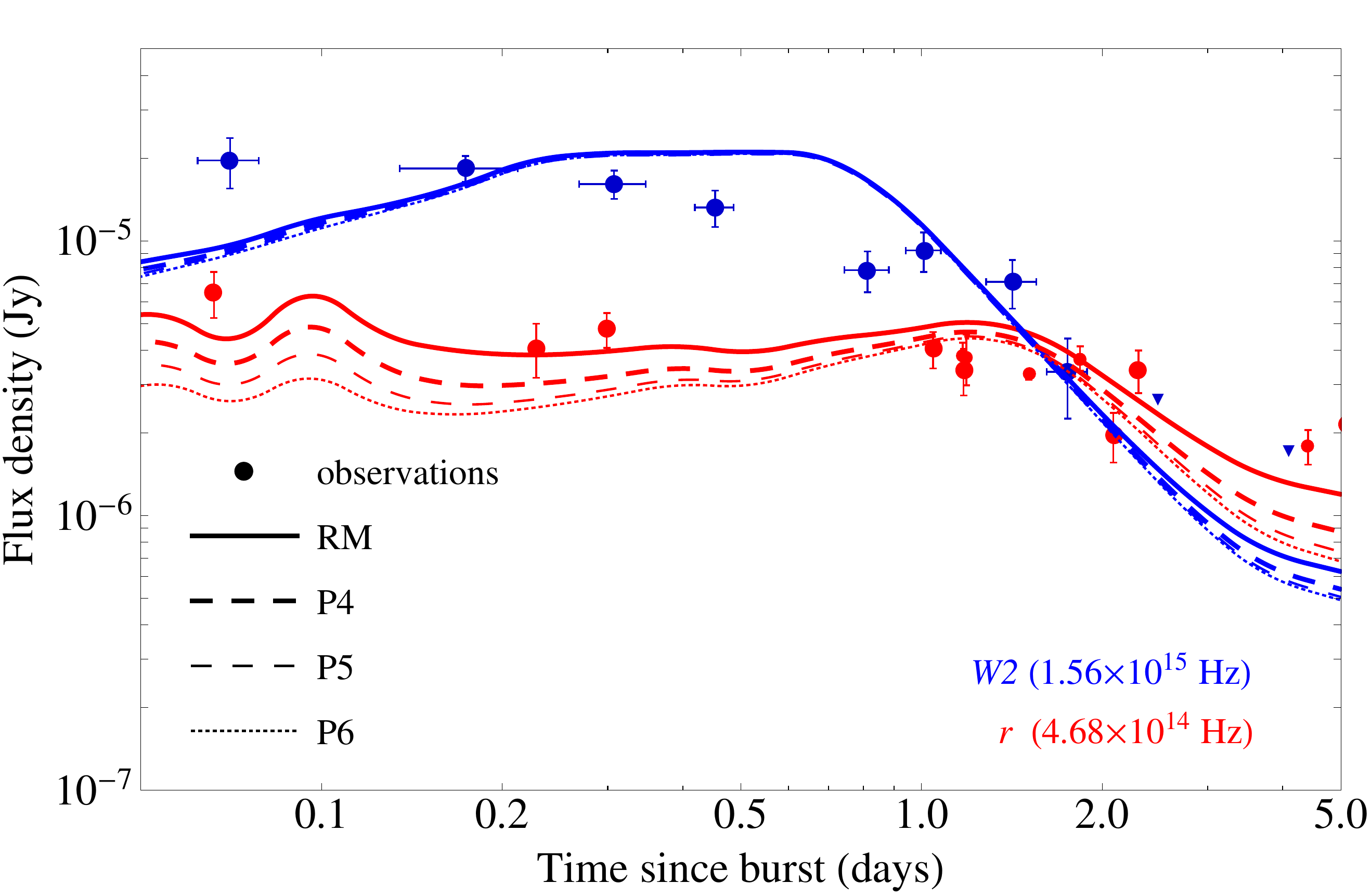}
\caption{Synthetic LCs for different values of the electron energy distribution index $p$:  $2.3$ (RM, solid lines), $2.4$ (P4, thick dashed lines), $2.5$ (P5, thin dashed lines) and $2.6$ (P6, dotted lines). Observational data are marked with symbols (see Fig.~\ref{fig:NTcontributions}).}
\label{fig:LCp}
\end{figure}

Fixing the hydrodynamical parameters used for the RM, we show how a change in the power-law index of the electron energy distribution affects the LCs in Fig. \ref{fig:LCp}. Values up to $p \approx 2.6$ are compatible with the first 5 d of observations in the $W2$ band. However, a value $p=2.3$ yields synthetic LCs closer to the observations in the $r$ band. The $W2$-band flux remains unchanged until $\simeq 2\,$d, independent of $p$, since in this band the thermal flux density is dominant at early times. At later times, smaller values of $p$ tend to produce a flatter and slightly more luminous LC, but still compatible with the observational upper limits available after $\sim 2$ d. 

\begin{figure}
\centering
\includegraphics[width=8.4cm]{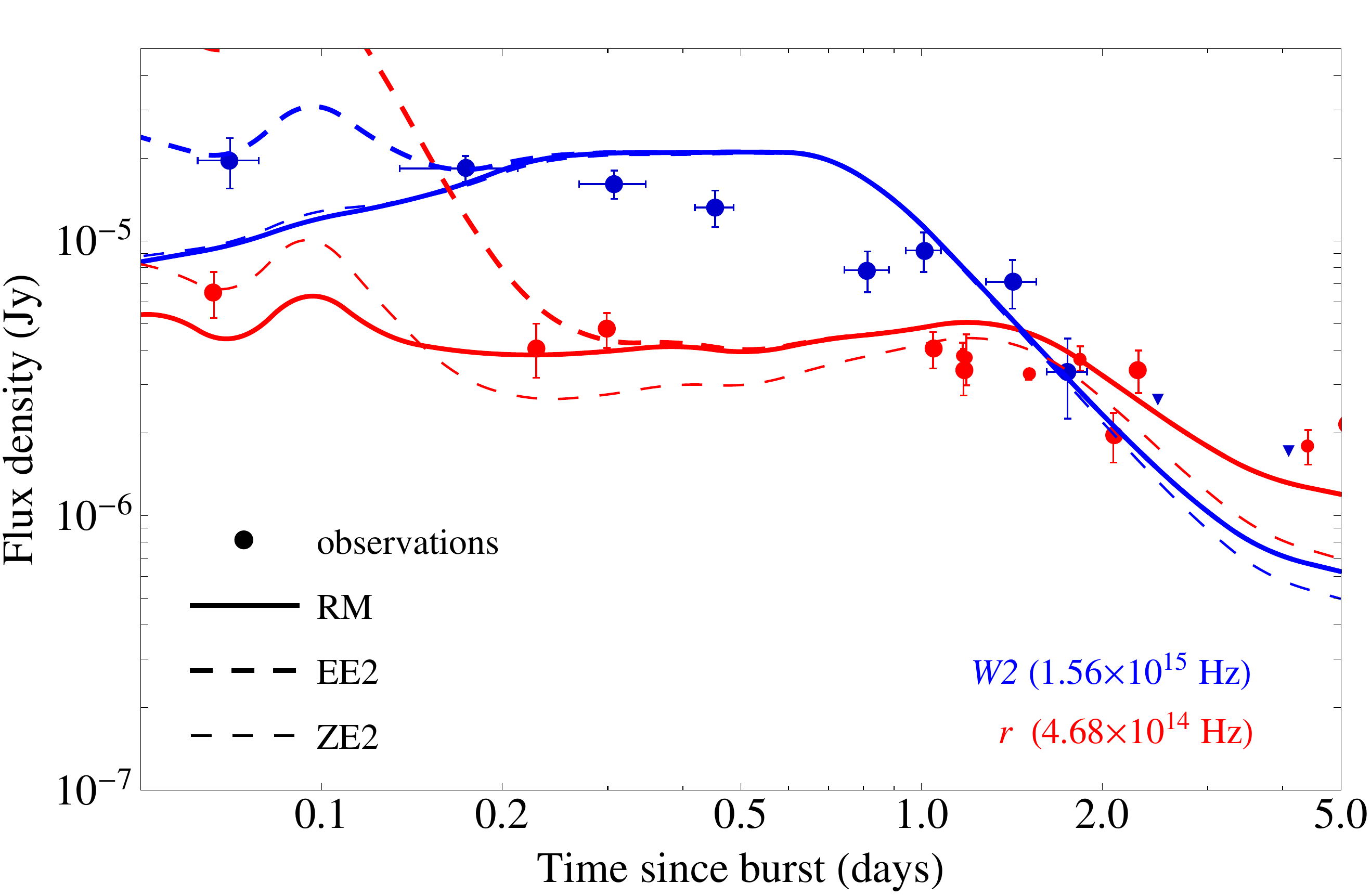}
\caption{Synthetic LCs for different values of $\epsilon_{\rm e}$ and $\zeta_{\rm e}$, respectively: $10^{-3}$ and $10^{-1}$ (RM, solid lines); $10^{-2}$ and $10^{-1}$ (EE2, thick dashed lines) and $10^{-3}$ and $10^{-2}$ (TE2, thin dashed lines). Observational data are marked with symbols (see Fig.~\ref{fig:NTcontributions}).}
\label{fig:LCe}
\end{figure}

Among the parameters used to model the synchrotron emission, $\epsilon_{\rm e}$ is the one which affects more the emission in both the $W2$ and in the $r$ (Fig.~\ref{fig:LCe}). We find that an increment of $\epsilon_{\rm e}$ is reflected in the LC as a flux increase, especially at low frequencies and early times, which is incompatible with observations (see model EE2; Table~\ref{tab:params}). A noticeable effect is that the non-thermal emission in the $W2$ band becomes the dominant contribution to the total flux until $\simeq 0.2\,$d, in contrast to the dominance of the thermal radiation in our RM (compare thick dashed and solid blue lines in Fig.~\ref{fig:LCe}). Decreasing $\zeta_{\rm e}$ by the same factor as $\epsilon_{\rm e}$ one obtains the same value for the effective fraction of the energy imparted to relativistic electrons, $\epsilon'_{\rm e}$, but the LCs are different. Decreasing $\zeta_{\rm e}$  (see model ZE2; Table~\ref{tab:params}) decreases the number of radiating electrons, yielding a higher flux at early times because there is more energy per electron. However, when the electrons have cooled at later times, the flux decreases compared to the reference case (which initially contains more electrons at lower energies).

\begin{figure}
\centering
\includegraphics[width=8.4cm]{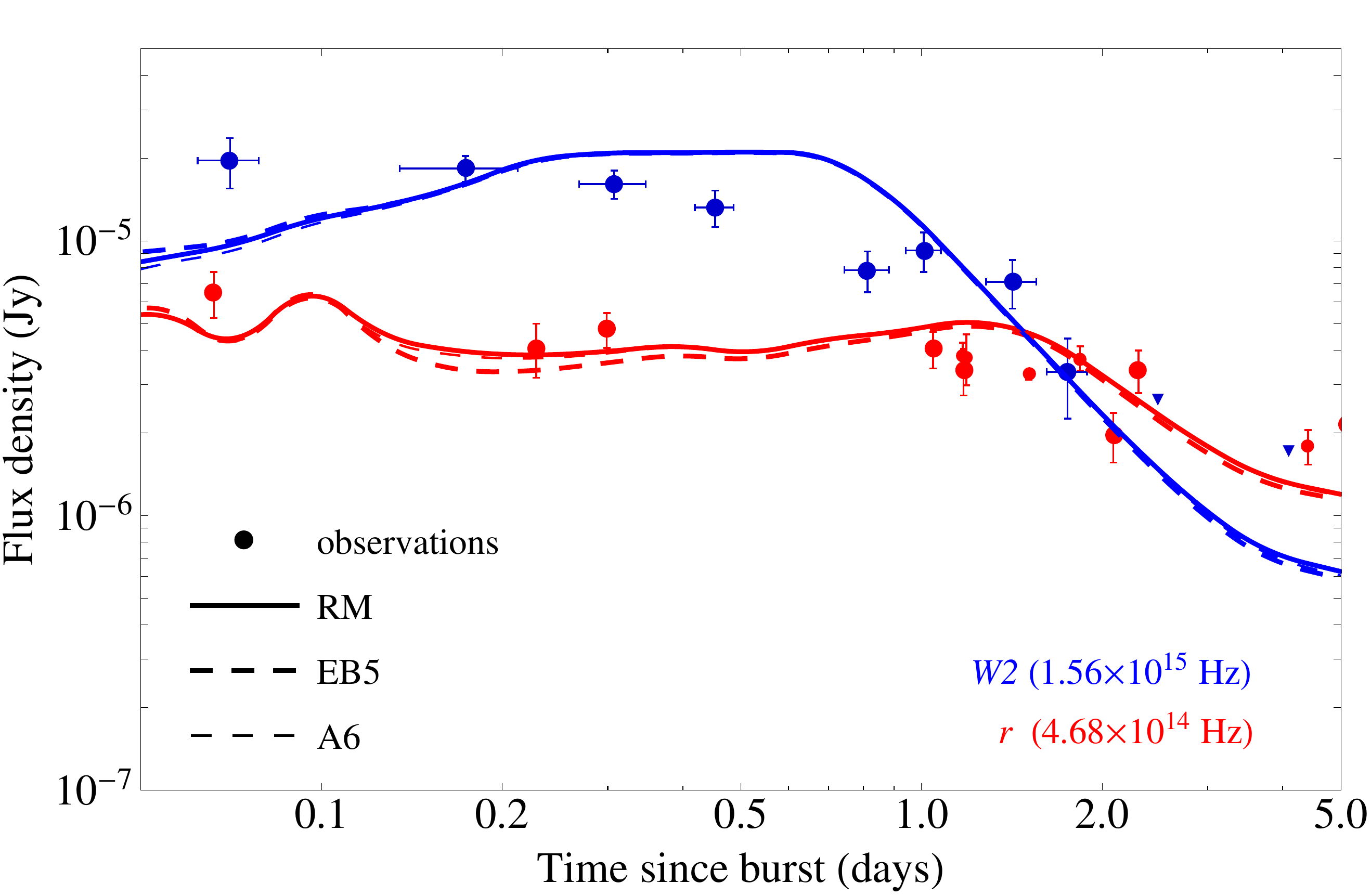}
\caption{Synthetic LCs for different values of $\epsilon_B$ and $a_{\rm acc}$, respectively: $10^{-6}$ and $1$ (RM, solid lines); $10^{-5}$ and $1$ (EB5, thick dashed lines); $10^{-6}$ and $10^6$ (A6, thick dashed lines). Observational data are marked with symbols (see Fig.~\ref{fig:NTcontributions}).}
\label{fig:LCB}
\end{figure}

In Fig.~\ref{fig:LCB} we show the effect of changing $\epsilon_B$ in the RM.  An increase by a factor of $10$ of $\epsilon_B$ (which is on reach of our models; see Sect.~\ref{sec:microparams}) has almost no influence on the observed flux. In the same Fig.~\ref{fig:LCB} we display the effect of changing $a_{\rm acc}$. It is evident that the variation of $\gamma_{\rm max}$ (equation~\ref{eq:gamma2}) of the particle distribution produced by the change in $a_{\rm acc}$ ($\gamma_{\rm max}\propto a_{\rm acc}^{-1/2}$) is insufficient to induce an observable variation on the global behaviour of the optical LCs. The reason for it is that the typical Lorentz factors contributing to the synchrotron emission in the $W2$ and $r$ bands ($\sim 9\times 10^3$ and $\sim 5\times 10^3$, respectively) are smaller than $\gamma_{\rm max}$ when $a_{\rm acc}$ is either 1 ($\gamma_{\rm max}\sim 2\times 10^7$) or $10^6$ ($\gamma_{\rm max}\sim 2\times 10^4$).

\subsubsection{Isotropic energy of the jet, $E_{\rm iso}$}

\begin{figure}
\centering
\includegraphics[width=8.4cm]{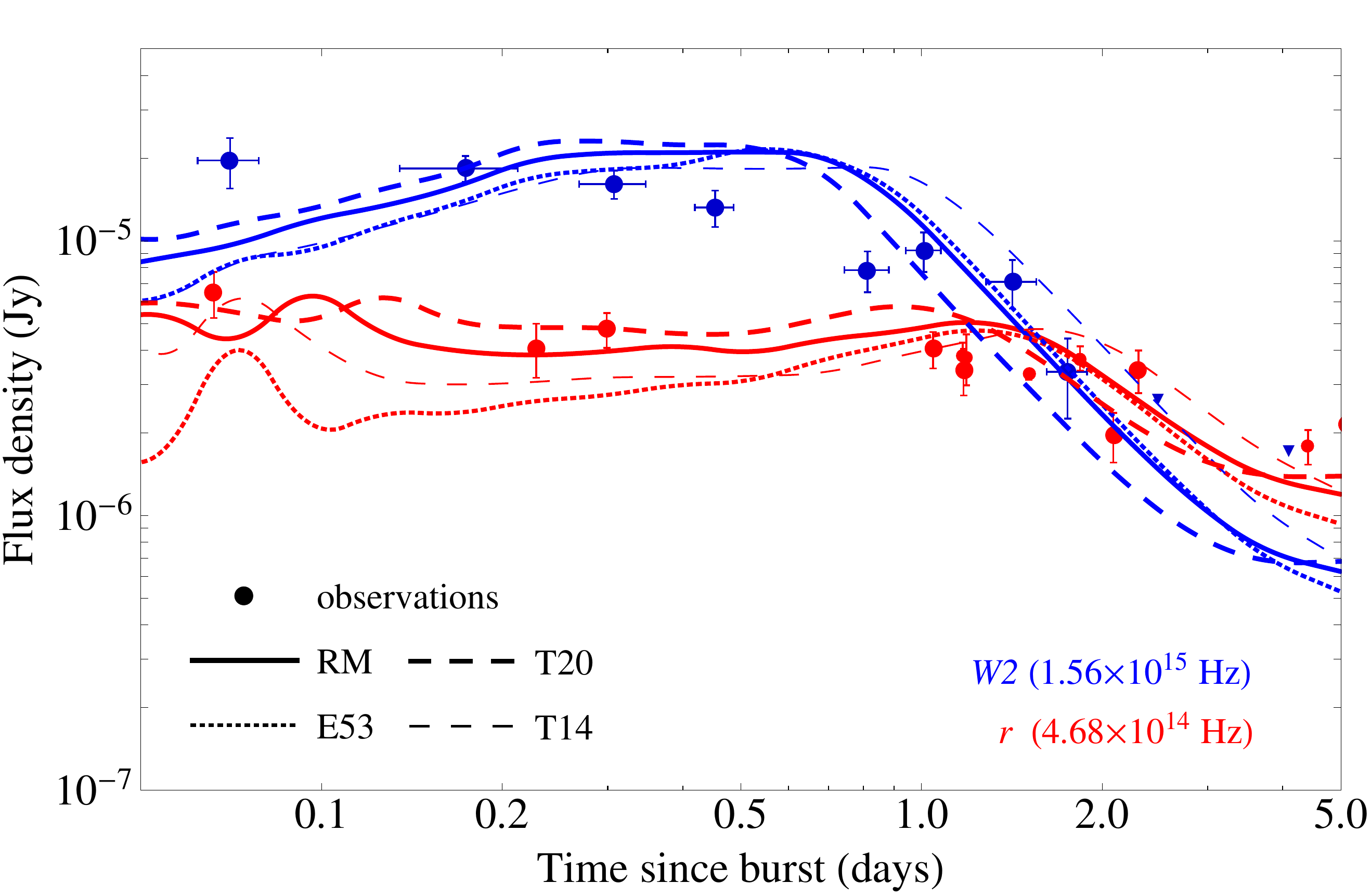}
\caption{Synthetic LCs for different jet energies. Models RM (solid lines), E53 (dotted lines), T20 (thick dashed lines) and T14 (thin dashed lines) are depicted. Observational data are marked with symbols (see Fig.~\ref{fig:NTcontributions}).}
\label{fig:LCEiso}
\end{figure}

One of the parameters under consideration is the equivalent isotropic energy of the jet, for which we only have a lower observational limit. We have considered two different models with isotropic energies of $4\times10^{53}$ (RM) and $2\times10^{53}$ (E53), keeping $\theta_{\rm j} = 17^\circ$ constant. The change in $E_{\rm iso}$ causes important differences associated with the emission of non-thermal radiation, while the thermal contribution, which is dominant from $\ga 0.2$\,d, is similar in both cases. More energetic jet models develop stronger bow shocks (i.e. stronger forward shocks) and, hence, the associated non-thermal emission is more powerful. In Fig.~\ref{fig:LCEiso} we see that for the least energetic jet model, E53, the flux at early times is low compared with observations. A slightly larger value of $\epsilon_{\rm e}$ could increase the early time flux, so that the data could still be fitted with a larger non-thermal component until $\lesssim 0.15\,$d. At later times it can be seen that the total emission is comparable to that of RM.

\subsubsection{Half-opening angle, $\theta_{\rm j}$}

For a fixed value of the equivalent isotropic energy of the jet, the true injected energy ($E_{\rm j}$) depends on the assumed half-opening angle of the jet. This means that for a constant energy per solid angle broader jets are more energetic: $E_{\rm j} = E_{\rm iso} (1 - \cos \theta_{\rm j}) /2$. In the model T14 the spectral inversion is produced about half a day later (thin dashed line in Fig.~\ref{fig:LCEiso}) than in RM. On the other hand, in the LC of T20 (thick dashed line in Fig.~\ref{fig:LCEiso}) the spectral inversion happens earlier than it does observationally (by almost half a day), and earlier than in the RM. After the spectral inversion in the model with the largest half-opening angle, the decay rate of the flux is very similar to the RM. Our conclusion is that values $\theta_{\rm j}\simeq 17^\circ$ are those which reproduce the observations better.
%

\subsubsection{CE-shell density contrast with respect to the external medium, $\rho_\CEsh/\rho_{\rm ext}$}
\label{sec:CEdensity}
Assuming a fixed CE-shell geometry, a change of a rest-mass density of the shell, $\rho_\CEsh$, yields an equivalent change in its total mass. We take moderate values for the CE-shell mass,  $M_\CEsh \sim 0.26$, $0.14 M_\odot$ (see Paper I). This corresponds to the rest-mass densities $\rho_\CEsh/\rho_{\ext} = 1500$ (RM), $817$ (D2), respectively, if the shell is uniform. We also consider model GS which incorporates a CE shell that has a rest-mass density decay as $r^{-2}$ (see Table~\ref{tab:params}). We have tuned the rest-mass density at $R_{\CEin}$ in order to have approximately the same mass in the CE shell of model GS as in the RM. In model GS the density jump across the inner edge of the CE shell is about three times larger than in the RM, while at the outer edge of the CE shell the density jump is halved with respect to the RM. The change in the distribution of rest mass of the CE shell yields small differences in the $W2$- and $r$-band LCs of model GS (see fig.~19 of Paper I). The observed flux in these two bands integrated up to the peak frequency in each band is $\gtrsim10$ per cent larger than in the RM, and the peak at each frequency is shifted to a bit earlier times. After the maxima, the decay of the LCs is slightly faster than in the RM. 

\begin{figure}
\centering
\includegraphics[width=8.4cm]{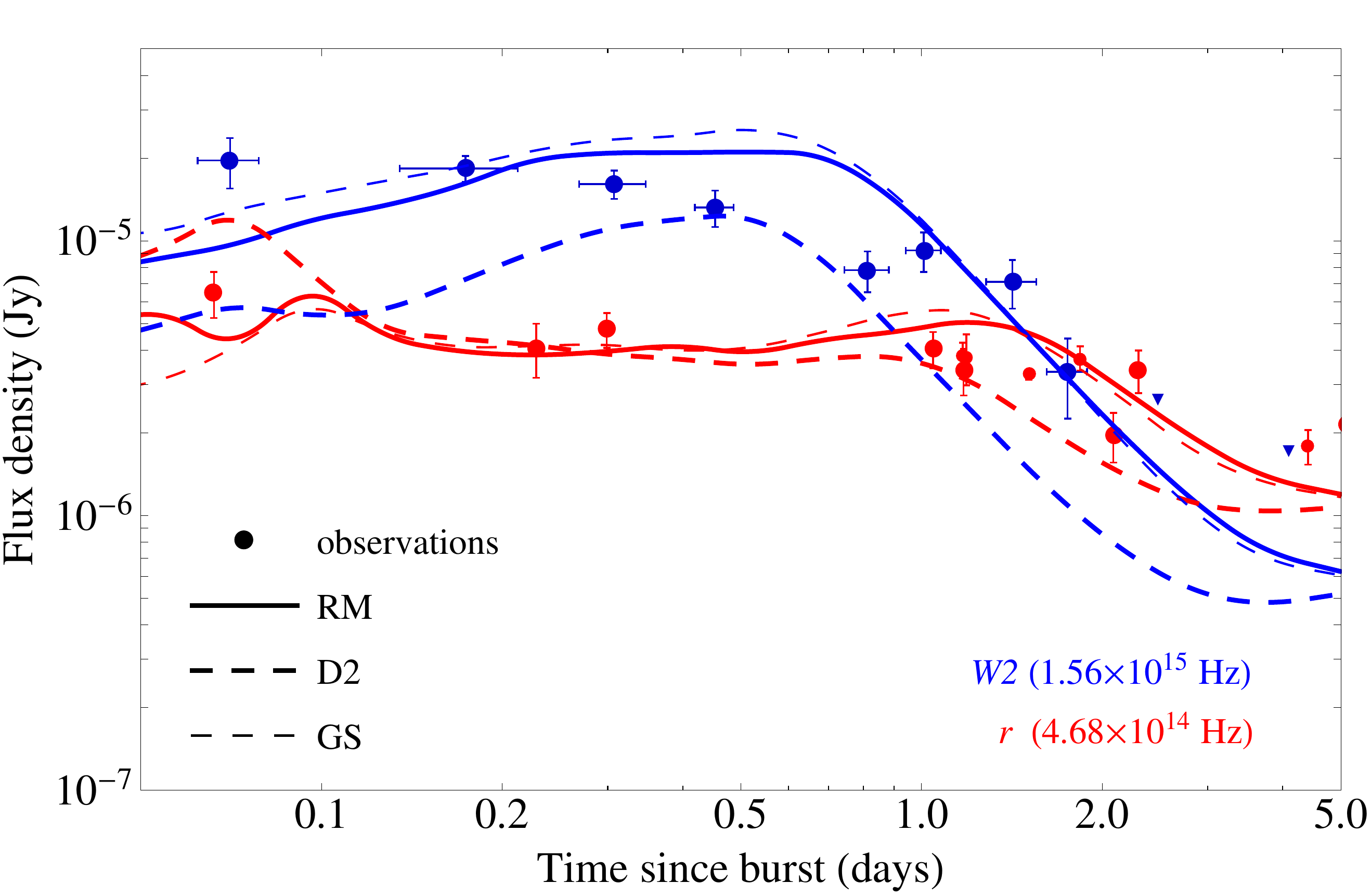}
\caption{Synthetic LCs obtained varying the rest-mass density,  $\rho_\CEsh$, and geometry of the CE shell:  RM (solid lines), D2 (thick dashed lines) and GS (thin dashed lines) models.  Observational data are marked with symbols (see Fig.~\ref{fig:NTcontributions}).}
\label{fig:LCdens}
\end{figure}

For model D2 the $W2$-band flux is halved with respect to that of the RM in the period under study (Fig.~\ref{fig:LCdens}, blue solid and thick dashed lines). The response in the $r$ band to changes in the CE-shell density is non-monotonic (Fig.~\ref{fig:LCdens}, red solid and thick dashed lines).  At early times ($t_{\rm obs}\la 0.15\,$d), when the flux is dominated by the non-thermal contribution,  the observed emission grows in model D2 by a factor of 2 with respect to the RM. Afterwards it is smaller and does not fit the data very well. Therefore, a lower shell density causes the thermal flux to decrease. The reason for this behaviour is that different CE-shell densities will cause differences in the jet/shell interaction. The lower density CE shell allows the jet to travel through the funnel more easily, dragging less mass and breaking out of the shell outer radius at a higher speed. During this epoch the jet FS is still rather relativistic and, consistently, an afterglow-like emission dominates the observed flux. The opposite happens for a more massive CE shell, where the jet is slowed down much more and the afterglow-like emission is more suppressed.

\subsubsection{CE-shell geometry}

As we have shown in Paper I, most of the observed thermal radiation comes from the interaction region between the CE shell and the jet. In previous sections, we have argued that the jet/CE-shell interaction also produces the elimination of the RS and, furthermore, depending on the details of such interaction, the emission associated with shocks in that region may also yield a significant change in the observed flux at early times. Therefore, the exact properties of the shell and the funnel geometry can significantly influence both the thermal and the non-thermal emission signatures of our models. 

\begin{figure}
\centering
\includegraphics[width=8.4cm]{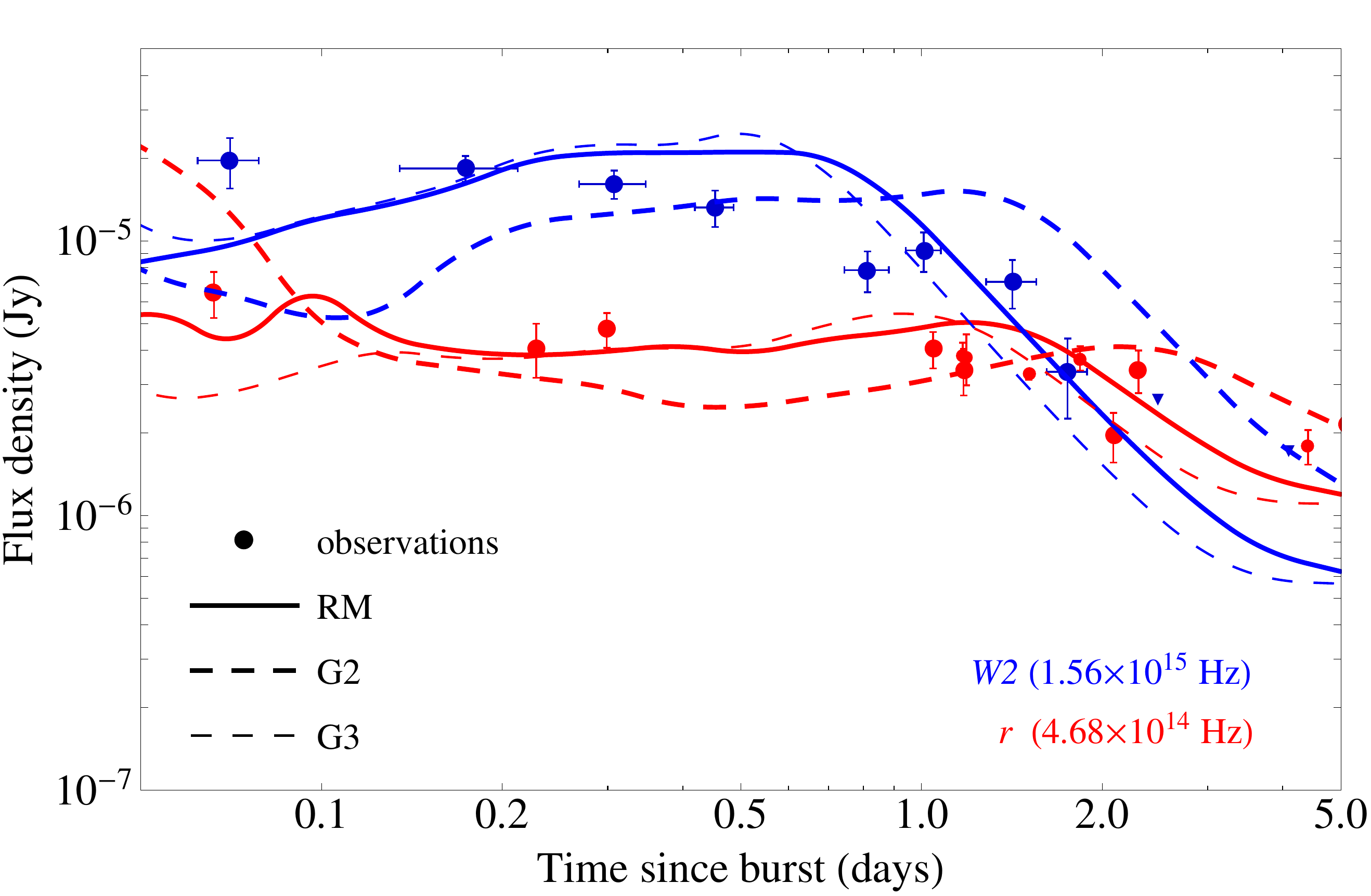}
\caption{Synthetic LCs obtained varying the rest-mass density and geometry of the CE shell, $\rho_\CEsh$:  RM (solid lines), G2 (thick dashed lines) and G3 (thin dashed lines) models.  Observational data are marked with symbols (see Fig.~\ref{fig:NTcontributions}).}
\label{fig:LCgeo}
\end{figure}

We have tested three different CE-shell funnel geometries: two toroidal ones (RM and G3) and a simpler, linear geometry (G2) for the funnel (see fig. 1 of Paper I). We use the same CE-shell density in these all three models. Therefore, the shell mass in the case of linear opening of the funnel (G2) is a bit smaller than in the RM, with a toroidal-like shell funnel, and this one is even smaller than in the model G3 with a more closed funnel. Fig.~\ref{fig:LCgeo} shows that a toroidally shaped CE-shell geometry explains a bit better the observations. As in the case of reducing the CE-shell density (Sect.~\ref{sec:CEdensity}), reducing the amount of mass that the jet sweeps in the angular region $\theta_{\rm f,in} \le \theta \le \theta_{\rm j}$ results in a reduced jet/shell interaction and less flux at early times in $W2$ band. The $r$ band is still dominated at early times by non-thermal radiation. The spectral inversion in G2 happens later than in RM due to the initially smaller jet/shell interaction, which delays the CE-shell effective ablation. This gives a later peak in $W2$ band and the flattening of the $r$ band up to 5 d. 

Model G3 is the one which has the larger mass of what we called {\em CE-early-interaction wedge} in Paper I. For this reason the jet/shell interaction is stronger initially (although non-thermal radiation is not dominant at the early stages of evolution) and the CE shell is ablated even sooner than in RM, giving an earlier peak of flux and an earlier redenning of the system (Fig.~\ref{fig:LCgeo}).

\subsubsection{External medium}

\begin{figure}
\centering
\includegraphics[width=8.4cm]{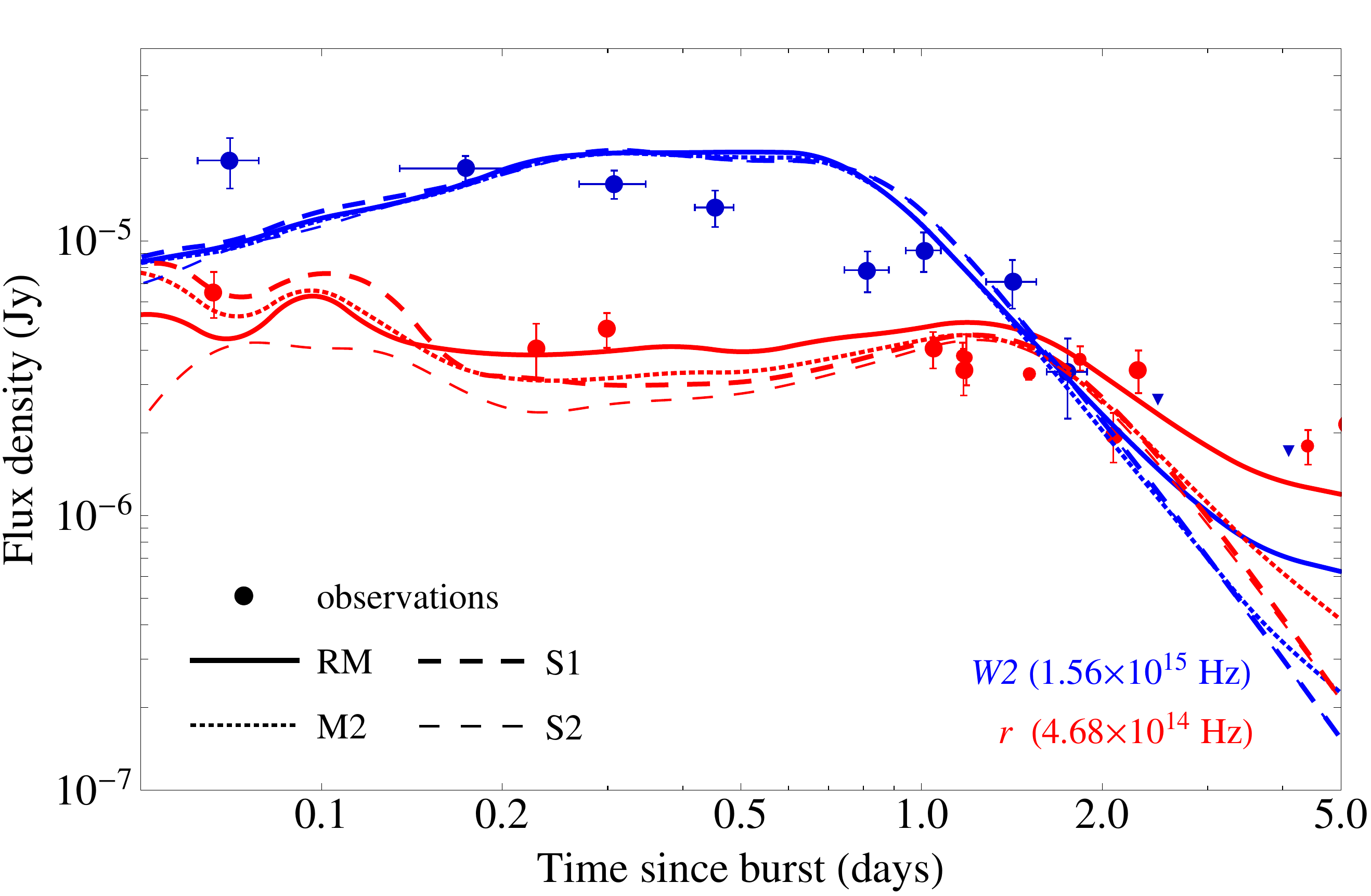}
\caption{Synthetic LCs for different external media. We show two models with uniform media, RM (solid lines) and M2 (dotted lines), as well as two stratified media S1 (thick dashed lines) and S2 (thin dashed lines). The stratification of the EM is set in the in rest-mass density and pressure as $\propto r^{-1}$ (S1) and $\propto r^{-2}$ (S2). Observational data are marked with symbols (see Fig.~\ref{fig:NTcontributions}).}
\label{fig:LCextmedium}
\end{figure}

As we seen in Paper I, the thermal emission does not depend on the EM structure. It is mainly produced by the jet/CE-shell interaction. However at later times ($t_{\rm obs} \ga 3$ d), after the CE shell has been completely ablated, a flattening in the LC can be observed. This emission is associated with the thermal contribution of the bubble blown by the jet, whose expansion rate depends chiefly on the mass plowed by the FS from the EM. This emission is expected to be reduced in models with a low-density EM (e.g., in model M2). In our models with a rest-mass density (and pressure) declining external media (S1 and S2) the mass of EM swept by the FS after the same time after the jet injection is smaller, singularly after a lab-frame time $T\gtrsim 0.6\,$d (see fig.~4 of Paper I). Thus, the average rest mass of the cavity in models S1 and S2 is 5--10 times smaller than in the RM and, thereby, the thermal bremsstrahlung emissivity (proportional to the square of the rest-mass density) is much smaller. The consequence is that the synthetic LCs of models S1 and S2 do not tend to flatten after $t_{\rm obs}\gtrsim 3\,$d, but decay with time as $\sim t_{\rm obs}^{-2.7}$ (Fig.~\ref{fig:LCextmedium}). 

The way in which the stratification of the EM affects the non-thermal emission is more complex. At late times, only the FS is still able to efficiently accelerate non-thermal electrons. The fraction of energy transferred to the electrons depends on the density of the external shock where they are injected. This density mostly depends on the EM conditions, so that at smaller density of the EM, one obtains a smaller density of the region shocked by the external shock, and consistently a smaller non-thermal flux density. All this causes the total emission at times $t_{\rm obs}\gtrsim 0.2\,$d to be dominated by the thermal emission, rather than by the FS. In the RM this situation may change after $t_{\rm obs}\gtrsim 5\,$d, when the FS flux density becomes comparable to that of the expanding shell in the UVOIR bands (Fig.~\ref{fig:NTcontributions}). This is not the case of model M2, in which a 10 times lower density in the EM results in a three times smaller flux than in the RM after 5\,d (Fig.~\ref{fig:LCextmedium}). At early times,  the thermal emission flux density is basically the same in stratified and non-stratified models, since the CE shell is uniform in both cases, and the thermal emission is linked to the jet/CE-shell interaction (see fig.~13 of Paper I). The non-thermal emission shows a rather different early evolution depending on the EM rest mass and pressure gradients. Our models S1 and S2 have a fiducial pressure, $p_\ext$, 100 times larger than in the RM. While the pressure quickly decreases below that of the RM in model S2, the linear pressure decrease of model S1 makes that the environmental pressure in this model equals that of the uniform medium of the RM only a long time after the jet head has left the outer radial CE-shell boundary. Before this time, the jet of model S1 encounters an EM with a (much) larger pressure than that of the RM or the S2 models. In particular, this happens while the jet is still advancing along the funnel. As a consequence, the strength of both the FS and the RS is larger in model S1 than in models RM and S2 and, hence, the non-thermal flux emitted by this model is larger. Conversely, the pressure along the funnel in model S2 is the smallest, thus the jet in this model develops the weaker (in relative terms) pair of FS and RS, resulting in the weakest early-time flux density of all three models.

The difference in strength of the forward and reverse shocks among models with different stratification of the EM also sets the times when the thermal flux becomes the dominant contribution to the observed emission. For model S2, the non-thermal flux density is always smaller than the thermal one, except at about $t_{\rm obs}\sim 0.1\,$d, when they are comparable. For models RM, S1 and M2 the non-thermal flux density is larger than the thermal one until $t_{\rm obs}\sim 0.15\,$d (see Fig.~\ref{fig:LCextmedium} and compare with fig.~14 of Paper I). Furthermore, it is worth to notice that the $r$-band flux of RM lies below of that of M2 and S1 models, in spite of the fact that the latter models a less dense EM and, consistently, the number density of non-thermal particles in the FS is smaller. The larger flux then results from a larger Doppler boosting of the FS radiation, produced because the jet propagates faster in a lower density EM. In model S2, the initial extra Doppler boosting does not compensate the deficit in the emission which results from the lowest number density among all models at hand. After the initial phase, the thermal emission quickly becomes dominant since the jet advances in a decreasing rest-mass density EM.

\section{Discussion and conclusions}
\label{sec:conclusions}


In this paper we provide a robust set of numerical models aiming to explain the bizarre phenomenology of GRB 101225A (in particular) and of the so-called BBD-GRBs (in general). In Paper I we have focused on the details of the relativistic hydrodynamic simulations supporting the model of T11, as well as explaining the origin of the thermal radiation. According to our model, the thermal signature in BBD-GRBs should be attributed to the interaction of an ultrarelativistic jet with a dense hydrogen shell in the vicinity of the progenitor star. Such a shell may result from the merger of a neutron star and the helium core of an evolved, massive star after a CE evolution. We have also shown in Paper I that the lack of a classical afterglow (or the presence of a very faint one) also results from the jet/CE-shell interaction, which baryon loads the beam of the jet leading to a much faster deceleration than what one would expect in the case in which the EM of the GRB progenitor star would not harbour such an `obstacle' for the jet propagation. 

Since the CE shell is cold and much denser than the jet, the jet/CE-shell interaction stimulates the thermal bremsstrahlung emission of the shocked CE shell, and we receive a long-lasting thermal emission that decays very quickly once the CE shell is fully disrupted by the jet and the system becomes transparent (see Paper I). The time it takes to disrupt the CE shell depends on the balance of several factors. Less massive shells (e.g. model D2) are more rapidly ablated and show a clear flux density deficit with respect to the observations and with respect to the RM. However, it is also important how the mass is distributed in the shell. Shells that are more massive closer to the axis of the system (e.g. our model G3, where the funnel of the toroidal shell is narrower than in the RM) or wider jets (e.g. model T20, with a larger half-opening angle than the RM) tend to increase the strength of the jet/CE-shell interaction and they can also be disrupted faster. Since our models assume a geometry and a mass for the CE shell (rather than computing the CE-shell geometry as a result of the neutron star/He star merger), we can only qualitatively compare our results with the observations. 

Associated to the effective ablation of the CE shell we find a reddening of the observed radiation (see also Paper I). A short time before the system becomes optically thin, the $W2$-band flux density starts to decay much faster than that in the $r$ band, eventually producing a spectral inversion that is also noticeable in the observations. Indeed, model RM captures very precisely this transition and that has been one of the main reasons to choose it as a reference case.

Though it is true that we do not observe a classical afterglow signature, our jet models still develop a hot cavity bounded by a pair of forward and reverse shocks, whose synchrotron emission has not been included in our previous work. Here, we have made a proper comparison between the numerical models and the first 5 d of UVOIR observations post-processing the relativistic hydrodynamic models of Paper I with a full transport scheme considering not only thermal emission processes (free--free bremsstrahlung), but also non-thermal (synchrotron) emission. We have found in Paper I that a major fraction of the observed flux during the first 5 d of the UVOIR observations can be attributed to the thermal emission of the jet/CE-shell interaction. However, in the very early stages of the event ($t_{\rm obs}\lesssim 0.2\,$d) we pointed out that we had a deficit of flux in the UVOIR bands, as well as in the X-ray band. In this paper we show that the forward shock synchrotron emission is very important to compensate for the thermal flux deficit at early times, particularly at low optical frequencies ($r$ band). The contribution of the non-thermal radiation in the X-ray band is very small and can be neglected for practical purposes. The emission of the RS is comparable to that of the FS during, approximately, the first $\sim 80$\,min of evolution. After that, it quickly decays because the RS is smeared out as a result of the jet/CE-shell interaction. This interaction generates a number of shocks which move almost perpendicularly to the jet propagation direction. Despite adding a minor contribution to the observed non-thermal flux density, such shocks cross the jet beam and eventually reach the jet axis. Along the way, they raise the rest mass and the entropy of the ultrarelativistic beam with the side effect that the jump in the hydrodynamical variables through the RS is basically cancelled. 

The quantitative contribution of non-thermal (synchrotron) processes to the observed flux depends on the set of microphysical parameters adopted. In the present paper, we have assumed somewhat low effective efficiencies for the RM, the one which accommodates better the observations.  For the RM, having an isotropic equivalent energy $E_{\rm iso}=4\times 10^{53}\,$erg, values of $\epsilon_{\rm e}'=\epsilon_{\rm e}/\zeta_{\rm e}\lesssim 0.1$ and $\epsilon_B=10^{-5}$--$10^{-6}$ are needed to not overpredict the UVOIR flux at early times. Larger efficiencies $\epsilon_{\rm e}'\simeq 0.1$ and $\epsilon_B\simeq 10^{-3}$ are also marginally compatible with observations if $E_{\rm iso}\simeq 10^{53}\,$erg (though the $W2$ band displays an obvious flux deficit in this case). We point out that technical difficulties prevent raising $\epsilon_B$ in our models. Values of $\epsilon_B$ in excess of $10^{-4}$ yield synchrotron cooling time-scales which are smaller than the typical time steps of our hydrodynamical models and cause order of magnitude errors in the estimation of the non-thermal flux density (particularly at high optical or X-ray frequencies). 

In the present paper, we have restricted the analysis to the case in which the microphysical parameters are uniform and time independent for all shocks. However,  it seems quite plausible that the RS has a larger value of $\epsilon_B$ than the FS \citep[see e.g.][]{DePasquale_etal_2013arXiv1312.1648}. Should that be the case, the RS signature might be more prominent during the first $\sim 0.07\,$d. Values of the power-law index, of the energy injection spectrum of the electrons, $p$, in excess of the reference value ($p=2.3$), tend to reduce a bit the total flux density during the first $\simeq 0.6\,$d of evolution, making them less likely, unless we combine an increase of $p$ with a moderate increase in e.g. $\epsilon_{\rm e}$. 

Since the dominant thermal contribution in our models is not associated with the expanding cavity blown by the jet, and since the non-thermal emission produced by the FS (that strongly depends on the properties of the EM) is subdominant throughout most of the evolution, the exact structure of the EM does not show up until observing times $t_{\rm obs}\gtrsim 2\,$d. By that time, the flux density decreases more shallowly in models with a uniform and high-density EM than in models with a density decaying or uniform low-density ones. In the uniform, high-density medium, at low optical frequencies ($r$ band), the flux decays as $\propto t_{\rm obs}^{-1.4}$ between the second and the third day, while it does as $\propto t_{\rm obs}^{-1.9}$ in the $W2$ band. This trend should be compared to the decay $\propto t_{\rm obs}^{-2.7}$ in both S1 and S2 models where the rest-mass density and pressure decrease as $r^{-1}$ and $r^{-2}$, respectively. Model M2 (uniform, low-density medium) shows a flux decay which lies between RM and models S1--S2. Remarkably, the time when the system becomes optically thin is also $t_{\rm obs}\simeq 1.5$--$2\,$d. The transition to transparency depends only on the time it takes the jet to completely ablate the CE shell, which is largely independent of the EM if the CE-shell mass is fixed and its radial extension is not drastically changed. The differences in the emission at early times caused by the EM structure are much more dependent on the parameters of the set-up than e.g. the time of transition to transparency and the typical location of the peak in different UVOIR bands. 

We point out that using the same radial gradients as in our models S1 and S2, but a (much) larger initial rest-mass density, so that the Sedov length of models S1 and S2 was the same as that of the RM (i.e. similar to the set-up of \citealt{DeColle_etal_2012ApJ...751...57}) may have brought a stronger non-thermal emission. However, our choice for a reference model having a uniform environment is motivated by the fact that under the assumption that the progenitor system of BBD-GRBs were neutron star/He star mergers, the environment of the later may not develop the strong winds typical of massive stars with solar metallicity. 

In a consistent simulation of the merger phase of a CE-binary system, the ejected CE shell will not have the sharp boundaries we are employing in our models to simplify the problem. Instead, we expect a relatively smooth transition between the He-core and the densest part of the CE shell. If the amount of mass in the region below the radius that we have named $R_{\rm CE,in}$ and the He-core is small compared with the mass between $R_{\rm CE,in}$ and  $R_{\rm CE,out}$, there will not be appreciable differences in the overall thermal emission with respect to the models in this paper (neither in the overall total emission). More notable differences may arise if the CE shell extends towards (much) larger radii, since in this case, if we fix the CE-shell mass (and consistently, reduce the average shell density), the emitted thermal flux (proportional to $\rho_{\rm CE,sh}^2$) will be strongly suppressed. Furthermore, the spectral reddening of the models will happen at a different time than in our RM. For instance, in the case in which $R_{\rm CE,out}$ is larger than assumed in the RM, but the CE-shell mass is fixed, the spectral inversion will happen earlier than in our RM. Hence, to reproduce the spectral reddening at the right time after the burst, our models favour CE ejecta that do not travel much further away than $\sim 10^{14}\,$cm during the final inspiral in of the neutron star on to the He-core, if the mass of the CE shell is a fraction of a solar mass.
\begin{figure}
\centering
\includegraphics[width=8.4cm]{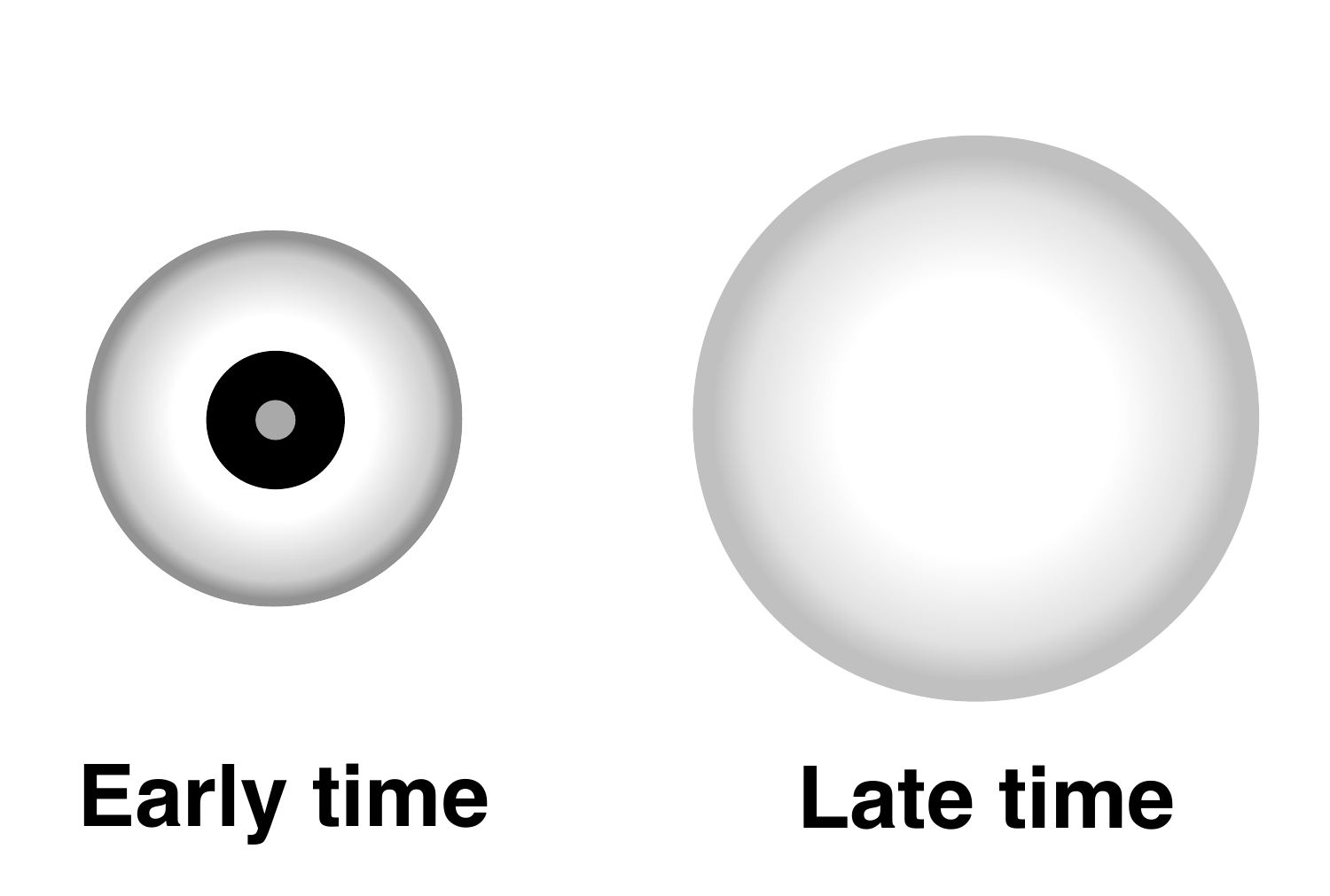}
\caption{Schematic representation of the distribution of received radiation at early times (left) and at late times (right) for an observer viewing the event at $0^\circ$. The grey-scale is roughly associated with the flux density, black being the largest flux and lighter grey shades showing lower intensity values. The figure is not directly obtained from simulation data, though it is an attempt to sum up the general properties of most of the numerical models.}
\label{fig:emission_distribution}
\end{figure}

For an observer looking head on at the event, the spatial distribution of the received radiation will evolve with time (Fig.~\ref{fig:emission_distribution}). Initially, the thermal emission from the jet/CE-shell interaction region together with the non-thermal emission from the RS and the shocks resulting from the jet/CE-shell interaction will be concentrated around the centre of the observed region. In addition to these components, a comparable emission arising from the FS will be seen as limb brightened (Fig.~\ref{fig:emission_distribution}, left). At these early times, the emission pattern will be `core dominated'. As the CE shell is ablated by the jet, the core emission will weaken and, eventually, the residual thermal emission from the whole bubble and the non-thermal emission from the FS will be the only observable components, both of which will be limb brightened. Thus, at late times we expect a `limb-dominated' emission pattern. We predict that the change in time from core- to limb-dominated emission will most probably produce have a significant change in polarization.

Finally, we have to mention that our treatment of the emission has certain limitations. For instance, we do not include the Comptonization of the radiation flux though we plan to incorporate it in the near future. Also, our models would need a higher resolution to properly resolve the CE-shell funnel if the CE-shell innermost radius ($R_{\CEin}$) was smaller. The interest of such a set up is that in T11, the X-ray emission of the standing hotspot is attributed to the shocked funnel, with a transversal size of a couple of solar radii ($\sim 10^{11}\,$cm). In this paper the cross-section of the funnel is $\simeq 8\times 10^{11}\,$cm, limited by the resolution we can have in the broad parametric scan of models we have performed. Hence, the X-ray synthetic LCs, tough broadly compatible with observations, display a clear excess of flux after $\simeq 0.3\,$d. We will address this issue in the future with higher resolution models.

\section*{Acknowledgements}
We acknowledge the support from the European Research Council (grant CAMAP-259276), and the partial support of grants AYA2010-21097-C03-01, CSD2007-00050 and PROMETEO-2009-103. CC-M also acknowledges the support of ACIF/2013/278 fellowship, and the partial support of UV-INV-PREDOC13-110509 fellowship. We thankfully acknowledge the computer resources, technical expertise and assistance provided by the Servei de Informatica at the University of Valencia. The research of AdUP and CT is supported by the Spanish research project AYA2012-39362-C02-02 and by Ram\'on y Cajal fellowships. AdUP acknowledges support from the European Commission under the Marie Curie Career Integration Grant programme (FP7-PEOPLE-2012-CIG 322307).

\appendix
\section{Spectral evolution}
\label{sec:spev}


We postprocess the hydrodynamical models with the radiative transport code \tiny{SPEV}\normalsize. In order to compute both the thermal and non-thermal specific intensity, we integrate independently the radiation transport equation for both cases, properly accounting for the emissivity of thermal bremsstrahlung and synchrotron processes, as well as for the corresponding absorption processes. In the following we will use the superscripts `th' and `nt' to refer to either thermal or non-thermal properties. 
For the computation of the total specific intensity we should consider a unique absorption coefficient that includes bremsstrahlung absorption, synchrotron self-absorption (both from synchrotron photons and from thermal photons), as well as absorption of synchrotron photons by the baryonic medium. However at the UVOIR and X-ray frequencies under consideration here, synchrotron self-absorption is negligible \citep*{Mimica_etal_2010}. Hence, the total specific intensity can be computed as the sum of the thermal and non-thermal contributions, $I_\nu=I_\nu^{\rm th}+I_\nu^{\rm nt}$, each of which changes along a photon path parametrized by the parameter $s$, as 
\begin{eqnarray}
\frac{dI_\nu^{\rm nt}}{ds} &=& j_\nu^{\rm nt} - (\alpha_\nu^{\rm th} + \alpha_\nu^{\rm nt}) I_\nu^{\rm nt}, \\
\frac{dI_\nu^{\rm th}}{ds}& =& j_\nu^{\rm th} - \alpha_\nu^{\rm th} I_\nu^{\rm th},
\label{eq:radiativetransport}
\end{eqnarray}
where $j_\nu^{\rm nt}$ and $\alpha_\nu^{\rm nt}$ are the synchrotron emission and absorption coefficient (see \citealt{Mimica_etal_2009ApJ}), and $j_\nu^{\rm th}$ and $\alpha_\nu^{\rm th}$ are the free--free bremsstrahlung emission and absorption coefficients (see Paper I). 
We note that non-thermal radiation coming from shocks in the CE-shell/jet interaction region and places where $r < R_{\CEin}$ will be strongly absorbed by the CE shell, and thus $\alpha_\nu^{\rm th} \gg \alpha_\nu^{\rm nt}$. Likewise bremsstrahlung absorption is almost negligible in the FS and RS since the rest-mass density at those shocks is much smaller than at the CE shell ($\rho_{\rm shocks} \ll \rho_\CEsh$), and the temperature is, conversely, larger ($T_{\rm shocks}\gg T_\CEsh$).


\subsection{Non-thermal emission}
\label{sec:app.nonthermal}

For the initialization of the non-thermal particle population we closely follow  \cite{Mimica_Aloy_2012MNRAS.421.2635}. As it is commonly done \citep[e.g.,][]{BM_1996ApJ...461L..37,DM_1998MNRAS.296..275,Mimica_etal_2004A&A...418..947,BDD_2009A&A...498..677,BD_2010ApJ...711..445}, we assume that a fraction $\epsilon_{\rm e}$ of the dissipated kinetic energy at the forward shock is used to accelerate electrons in the vicinity of the shock front.

We assume that the energy spectrum of the injected relativistic particles is a power-law in the electron Lorentz factor $\gamma$, with a power-law index $p$:
\begin{equation}\label{eq:elinj}
\dsfrac{\mathrm{d}n'_{\rm inj}}{\mathrm{d}t'\ \mathrm{d}\gamma} = Q_0\gamma^{-p} H(\gamma;\ \gamma_{\rm min}, \gamma_{\rm max})\ ,
\end{equation}
where $n'_{\rm inj}$ is the number density of the injected electrons, $Q_0$ is a normalization factor and $\gamma_{\rm min}$ and $\gamma_{\rm max}$ are the lower and upper injection cut-offs (computed below), all measured in the shocked fluid rest frame. The step function is defined as usual by $H(x; a, b) = 1$ if $a\leq x\leq b$ and $0$ otherwise.  The reference value for the power-law index of the electron energy distribution is $p = 2.3$.

We assume that the upper cut-off for the electron injection ($\gamma_{\rm max}$) is obtained by assuming that the acceleration time-scale is proportional to the gyration time-scale. Then the maximum Lorentz factor is obtained by equating this time-scale to the cooling time-scale,
\begin{equation}\label{eq:gamma2}
  \gamma_{\rm max} = \left(\dsfrac{3m_{\rm e}^2 c^4}{4\pi a_{\rm acc} e^3 B_{\rm st}}\right)^{1/2}\, .
\end{equation}
where $m_{\rm e}$ and $e$ are the electron mass and charge, respectively, $a_{\rm acc}\geq 1$ is the acceleration efficiency parameter \citep{BD_2010ApJ...711..445} and $B_{\rm st}$ is a stochastic magnetic field created by the action of shocks. We typically take here $a_{\rm acc} = 1$. The stochastic magnetic field strength is assumed to be a fraction $\epsilon_B$ of the internal energy density of the shocked fluid $u_{\rm S}$:
 \begin{equation}\label{eq:Bmic}
  B_{\rm st} = \sqrt{8\pi \epsilon_B u_{\rm S}}\, .
\end{equation}

The lower Lorentz factor cut-off ($\gamma_{\rm min}$) is obtained numerically by assuming that the number of accelerated electrons is a fraction $\zeta_{\rm e}$ of the electrons accelerated into the power-law distribution, so that it holds (see \citealt{Mimica_Aloy_2012MNRAS.421.2635} for details)
\begin{equation}\label{eq:gamma1}
\dsfrac{\int_{\gamma_{\rm min}}^{\gamma_{\rm max}}\ \mathrm{d}\gamma\
  \gamma^{1-p}}{\int_{\gamma_{\rm min}}^{\gamma_{\rm max}}\ \mathrm{d}\gamma\
  \gamma^{-p}} =
\dsfrac{\epsilon_{\rm e}}{\zeta_{\rm e}}\dsfrac{u_{\rm S}}{\Gamma'_0n_0 m_{\rm e}
  c^2}\, ,
\end{equation}
where $\Gamma'_0$ is the bulk Lorentz factor of the initially unshocked medium measured in the frame of the contact discontinuity between the initial states (up- and down-stream the shock front), and $n_0$ is the number density of the initially unshocked plasma measured in the fluid rest frame. Finally, to be consistent with our approximations for the calculation of the synchrotron emission, we only consider as synchrotron emitting those electrons whose Lorentz factor $\gamma_{\rm min}\geq 2$. Below this threshold we assume that electrons are purely thermal and that they only contribute to the free--free bremsstrahlung emission. The maximum Lorentz factor of the electron distribution, $\gamma_{\rm max}$, is determined by $B_{\rm st}$ and $a_{\rm acc}$. The condition $\gamma_{\rm min} < \gamma_{\rm max}$ must be fulfilled. 


\bibliographystyle{mn2e}
\bibliography{paper-mnras}

\begin{thebibliography}{}
\makeatletter
\relax
\def\mn@urlcharsother{\let\do\@makeother \do\$\do\&\do\#\do\^\do\_\do\%\do\~}
\def\mn@doi{\begingroup\mn@urlcharsother \@ifnextchar[{\mn@doi@}{\mn@doi@[]}}
\def\mn@doi@[#1]#2{\def\@tempa{#1}\ifx\@tempa\@empty
  \href{http://dx.doi.org/#2}{doi:#2}\else \href{http://dx.doi.org/#2}{#1}\fi
  \endgroup}
\def\mn@eprint#1#2{\mn@eprint@#1:#2::\@nil}
\def\mn@eprint@arXiv#1{\href{http://arxiv.org/abs/#1}{{\tt arXiv:#1}}}
\def\mn@eprint@dblp#1{\href{http://dblp.uni-trier.de/rec/bibtex/#1.xml}{dblp:#1}}
\def\mn@eprint@#1:#2:#3:#4\@nil{\def\@tempa {#1}\def\@tempb {#2}\def\@tempc
  {#3}\ifx \@tempc \@empty \let\@tempc\@tempb \let\@tempb\@tempa \fi \ifx
  \@tempb \@empty \def\@tempb{arXiv}\fi \@ifundefined
  {mn@eprint@\@tempb}{\@tempb:\@tempc}{\expandafter \expandafter \csname
  mn@eprint@\@tempb\endcsname \expandafter{\@tempc}}}

\bibitem[\protect\citeauthoryear{{Aloy}, {Ib{\'a}{\~n}ez}, {Mart{\'{\i}}}  \&
  {M{\"u}ller}}{{Aloy} et~al.}{1999}]{Aloy_1999ApJS}
{Aloy} M.~A.,  {Ib{\'a}{\~n}ez} J.~M.,  {Mart{\'{\i}}} J.~M.,   {M{\"u}ller}
  E.,  1999, \mn@doi [\apjs] {10.1086/313214}, \href
  {http://adsabs.harvard.edu/abs/1999ApJS..122..151A} {122, 151}

\bibitem[\protect\citeauthoryear{{Aloy}, {M{\" u}ller}, {Ib{\' a}{\~ n}ez},
  {Mart{\'{\i}}}  \& {MacFadyen}}{{Aloy}
  et~al.}{2000}]{Aloy_etal_ApJL_2000__Collapsar}
{Aloy} M.~A.,  {M{\" u}ller} E.,  {Ib{\' a}{\~ n}ez} J.~M.,  {Mart{\'{\i}}}
  J.~M.,   {MacFadyen} A.,  2000, \apjl, \href
  {http://cdsads.u-strasbg.fr/cgi-bin/nph-bib_query?bibcode=2000ApJ...531L.119A&amp;db_key=AST}
  {531, L119}

\bibitem[\protect\citeauthoryear{{Aloy} et~al.,}{{Aloy}
  et~al.}{2013}]{Aloy_etal_2013ASPC..474...33}
{Aloy} M.~A.,  et~al., 2013, in {Pogorelov} N.~V.,  {Audit} E.,   {Zank} G.~P.,
   eds,  Astronomical Society of the Pacific Conference Series Vol. 474,
  Numerical Modeling of Space Plasma Flows (ASTRONUM2012). p.~33

\bibitem[\protect\citeauthoryear{{Beloborodov}}{{Beloborodov}}{2010}]{Beloborodov_2010MNRAS_407_1033B}
{Beloborodov} A.~M.,  2010, \mn@doi [\mnras]
  {10.1111/j.1365-2966.2010.16770.x}, \href
  {http://adsabs.harvard.edu/abs/2010MNRAS.407.1033B} {407, 1033}

\bibitem[\protect\citeauthoryear{{B{\"o}ttcher} \& {Dermer}}{{B{\"o}ttcher} \&
  {Dermer}}{2010}]{BD_2010ApJ...711..445}
{B{\"o}ttcher} M.,  {Dermer} C.~D.,  2010, \mn@doi [\apj]
  {10.1088/0004-637X/711/1/445}, \href
  {http://adsabs.harvard.edu/abs/2010ApJ...711..445B} {711, 445}

\bibitem[\protect\citeauthoryear{{Bo{\v s}njak}, {Daigne}  \& {Dubus}}{{Bo{\v
  s}njak} et~al.}{2009}]{BDD_2009A&A...498..677}
{Bo{\v s}njak} {\v Z}.,  {Daigne} F.,   {Dubus} G.,  2009, \mn@doi [\aap]
  {10.1051/0004-6361/200811375}, \href
  {http://adsabs.harvard.edu/abs/2009A%26A...498..677B} {498, 677}

\bibitem[\protect\citeauthoryear{{Bykov} \& {Meszaros}}{{Bykov} \&
  {Meszaros}}{1996}]{BM_1996ApJ...461L..37}
{Bykov} A.~M.,  {Meszaros} P.,  1996, \mn@doi [\apjl] {10.1086/309999}, \href
  {http://adsabs.harvard.edu/abs/1996ApJ...461L..37B} {461, L37}

\bibitem[\protect\citeauthoryear{{Campana} et~al.,}{{Campana}
  et~al.}{2006}]{Campana_etal_2006Natur_short}
{Campana} S.,  et~al., 2006, \mn@doi [\nat] {10.1038/nature04892}, \href
  {http://adsabs.harvard.edu/abs/2006Natur.442.1008C} {442, 1008}

\bibitem[\protect\citeauthoryear{{Cuesta-Mart{\'{\i}}nez}, {Aloy}  \&
  {Mimica}}{{Cuesta-Mart{\'{\i}}nez} et~al.}{2014}]{Cuesta_Mimica_Aloy_2014_P1}
{Cuesta-Mart{\'{\i}}nez} C.,  {Aloy} M.~A.,   {Mimica} P.,  2014, \mnras, \href
  {http://adsabs.harvard.edu/abs/2014arXiv1408.1305C} {(Paper I)}

\bibitem[\protect\citeauthoryear{{Daigne} \& {Mochkovitch}}{{Daigne} \&
  {Mochkovitch}}{1998}]{DM_1998MNRAS.296..275}
{Daigne} F.,  {Mochkovitch} R.,  1998, \mn@doi [\mnras]
  {10.1046/j.1365-8711.1998.01305.x}, \href
  {http://adsabs.harvard.edu/abs/1998MNRAS.296..275D} {296, 275}

\bibitem[\protect\citeauthoryear{{De Colle}, {Ramirez-Ruiz}, {Granot}  \&
  {Lopez-Camara}}{{De Colle} et~al.}{2012}]{DeColle_etal_2012ApJ...751...57}
{De Colle} F.,  {Ramirez-Ruiz} E.,  {Granot} J.,   {Lopez-Camara} D.,  2012,
  \mn@doi [\apj] {10.1088/0004-637X/751/1/57}, \href
  {http://adsabs.harvard.edu/abs/2012ApJ...751...57D} {751, 57}

\bibitem[\protect\citeauthoryear{{De Pasquale} et~al.,}{{De Pasquale}
  et~al.}{2013}]{DePasquale_etal_2013arXiv1312.1648}
{De Pasquale} M.,  et~al., 2013, ArXiv e-prints, \href
  {http://cdsads.u-strasbg.fr/abs/2013arXiv1312.1648D} {}

\bibitem[\protect\citeauthoryear{{Gendre} et~al.,}{{Gendre}
  et~al.}{2013}]{Gendre_etal_2013ApJ...766...30G}
{Gendre} B.,  et~al., 2013, \mn@doi [\apj] {10.1088/0004-637X/766/1/30}, \href
  {http://adsabs.harvard.edu/abs/2013ApJ...766...30G} {766, 30}

\bibitem[\protect\citeauthoryear{{Giannios} \& {Spruit}}{{Giannios} \&
  {Spruit}}{2007}]{GS_2007A&A...469....1G}
{Giannios} D.,  {Spruit} H.~C.,  2007, \mn@doi [\aap]
  {10.1051/0004-6361:20066739}, \href
  {http://adsabs.harvard.edu/abs/2007A%26A...469....1G} {469, 1}

\bibitem[\protect\citeauthoryear{{Hasco{\"e}t}, {Daigne}  \&
  {Mochkovitch}}{{Hasco{\"e}t} et~al.}{2013}]{Hascoet_etal_2013A&A...551A.124H}
{Hasco{\"e}t} R.,  {Daigne} F.,   {Mochkovitch} R.,  2013, \mn@doi [\aap]
  {10.1051/0004-6361/201220023}, \href
  {http://adsabs.harvard.edu/abs/2013A%26A...551A.124H} {551, A124}

\bibitem[\protect\citeauthoryear{{Kouveliotou}, {Meegan}, {Fishman}, {Bhat},
  {Briggs}, {Koshut}, {Paciesas}  \& {Pendleton}}{{Kouveliotou}
  et~al.}{1993}]{Kouveliotou_etal_1993}
{Kouveliotou} C.,  {Meegan} C.~A.,  {Fishman} G.~J.,  {Bhat} N.~P.,  {Briggs}
  M.~S.,  {Koshut} T.~M.,  {Paciesas} W.~S.,   {Pendleton} G.~N.,  1993,
  \mn@doi [\apjl] {10.1086/186969}, \href
  {http://adsabs.harvard.edu/abs/1993ApJ...413L.101K} {413, L101}

\bibitem[\protect\citeauthoryear{{Kumar}, {Hern{\'a}ndez}, {Bo{\v s}njak}  \&
  {Duran}}{{Kumar} et~al.}{2012}]{Kumar_etal_2012MNRAS.427L..40}
{Kumar} P.,  {Hern{\'a}ndez} R.~A.,  {Bo{\v s}njak} {\v Z}.,   {Duran} R.~B.,
  2012, \mn@doi [\mnras] {10.1111/j.1745-3933.2012.01341.x}, \href
  {http://adsabs.harvard.edu/abs/2012MNRAS.427L..40K} {427, L40}

\bibitem[\protect\citeauthoryear{{Lazzati}, {Morsony}  \& {Begelman}}{{Lazzati}
  et~al.}{2009}]{Lazzati_etal_2009ApJ_700L_47L}
{Lazzati} D.,  {Morsony} B.~J.,   {Begelman} M.~C.,  2009, \mn@doi [\apjl]
  {10.1088/0004-637X/700/1/L47}, \href
  {http://adsabs.harvard.edu/abs/2009ApJ...700L..47L} {700, L47}

\bibitem[\protect\citeauthoryear{{Levan} et~al.,}{{Levan}
  et~al.}{2014}]{Levan_etal_2014ApJ...781...13_short}
{Levan} A.~J.,  et~al., 2014, \mn@doi [\apj] {10.1088/0004-637X/781/1/13},
  \href {http://adsabs.harvard.edu/abs/2014ApJ...781...13L} {781, 13}

\bibitem[\protect\citeauthoryear{{L{\'o}pez-C{\'a}mara}, {Morsony}, {Begelman}
  \& {Lazzati}}{{L{\'o}pez-C{\'a}mara}
  et~al.}{2013}]{Lopez_Camara_etal_2013ApJ...767...19L}
{L{\'o}pez-C{\'a}mara} D.,  {Morsony} B.~J.,  {Begelman} M.~C.,   {Lazzati} D.,
   2013, \mn@doi [\apj] {10.1088/0004-637X/767/1/19}, \href
  {http://adsabs.harvard.edu/abs/2013ApJ...767...19L} {767, 19}

\bibitem[\protect\citeauthoryear{{Lyutikov} \& {Blandford}}{{Lyutikov} \&
  {Blandford}}{2003}]{LB_2003astro.ph.12347L}
{Lyutikov} M.,  {Blandford} R.,  2003, ArXiv Astrophysics e-prints, \href
  {http://adsabs.harvard.edu/abs/2003astro.ph.12347L} {}

\bibitem[\protect\citeauthoryear{{Medvedev} \& {Loeb}}{{Medvedev} \&
  {Loeb}}{1999}]{ML_1999ApJ...526..697}
{Medvedev} M.~V.,  {Loeb} A.,  1999, \mn@doi [\apj] {10.1086/308038}, \href
  {http://adsabs.harvard.edu/abs/1999ApJ...526..697M} {526, 697}

\bibitem[\protect\citeauthoryear{{Meszaros} \& {Rees}}{{Meszaros} \&
  {Rees}}{1997}]{MR_1997ApJ}
{Meszaros} P.,  {Rees} M.~J.,  1997, \mn@doi [\apjl] {10.1086/310692}, \href
  {http://adsabs.harvard.edu/abs/1997ApJ...482L..29M} {482, L29}

\bibitem[\protect\citeauthoryear{{Mimica} \& {Aloy}}{{Mimica} \&
  {Aloy}}{2012}]{Mimica_Aloy_2012MNRAS.421.2635}
{Mimica} P.,  {Aloy} M.~A.,  2012, \mn@doi [\mnras]
  {10.1111/j.1365-2966.2012.20495.x}, \href
  {http://cdsads.u-strasbg.fr/abs/2012MNRAS.421.2635M} {421, 2635}

\bibitem[\protect\citeauthoryear{{Mimica}, {Aloy}, {M{\"u}ller}  \&
  {Brinkmann}}{{Mimica} et~al.}{2004}]{Mimica_etal_2004A&A...418..947}
{Mimica} P.,  {Aloy} M.~A.,  {M{\"u}ller} E.,   {Brinkmann} W.,  2004, \mn@doi
  [\aap] {10.1051/0004-6361:20034261}, \href
  {http://adsabs.harvard.edu/abs/2004A%26A...418..947M} {418, 947}

\bibitem[\protect\citeauthoryear{{Mimica}, {Aloy}, {M{\"u}ller}  \&
  {Brinkmann}}{{Mimica} et~al.}{2005}]{Mimica_etal_2005A&A...441..103}
{Mimica} P.,  {Aloy} M.~A.,  {M{\"u}ller} E.,   {Brinkmann} W.,  2005, \mn@doi
  [\aap] {10.1051/0004-6361:20053218}, \href
  {http://adsabs.harvard.edu/abs/2005A%26A...441..103M} {441, 103}

\bibitem[\protect\citeauthoryear{{Mimica}, {Aloy}  \& {M{\"u}ller}}{{Mimica}
  et~al.}{2007}]{Mimica_etal_2007A&A...466...93}
{Mimica} P.,  {Aloy} M.~A.,   {M{\"u}ller} E.,  2007, \mn@doi [\aap]
  {10.1051/0004-6361:20066811}, \href
  {http://adsabs.harvard.edu/abs/2007A%26A...466...93M} {466, 93}

\bibitem[\protect\citeauthoryear{{Mimica}, {Aloy}, {Agudo}, {Mart{\'{\i}}},
  {G{\'o}mez}  \& {Miralles}}{{Mimica} et~al.}{2009}]{Mimica_etal_2009ApJ}
{Mimica} P.,  {Aloy} M.-A.,  {Agudo} I.,  {Mart{\'{\i}}} J.~M.,  {G{\'o}mez}
  J.~L.,   {Miralles} J.~A.,  2009, \mn@doi [\apj]
  {10.1088/0004-637X/696/2/1142}, \href
  {http://adsabs.harvard.edu/abs/2009ApJ...696.1142M} {696, 1142}

\bibitem[\protect\citeauthoryear{{Mimica}, {Giannios}  \& {Aloy}}{{Mimica}
  et~al.}{2010}]{Mimica_etal_2010}
{Mimica} P.,  {Giannios} D.,   {Aloy} M.~A.,  2010, \mn@doi [\mnras]
  {10.1111/j.1365-2966.2010.17071.x}, \href
  {http://adsabs.harvard.edu/abs/2010MNRAS.407.2501M} {407, 2501}

\bibitem[\protect\citeauthoryear{Mizuta \& Aloy}{Mizuta \&
  Aloy}{2009}]{Mizuta_Aloy_2009}
Mizuta A.,  Aloy M.~A.,  2009, The Astrophysical Journal, 699, 1261

\bibitem[\protect\citeauthoryear{{Mizuta}, {Yamasaki}, {Nagataki}  \&
  {Mineshige}}{{Mizuta} et~al.}{2006}]{Mizuta_etal_2006ApJ...651..960M}
{Mizuta} A.,  {Yamasaki} T.,  {Nagataki} S.,   {Mineshige} S.,  2006, \mn@doi
  [\apj] {10.1086/507861}, \href
  {http://adsabs.harvard.edu/abs/2006ApJ...651..960M} {651, 960}

\bibitem[\protect\citeauthoryear{{Mizuta}, {Nagataki}  \& {Aoi}}{{Mizuta}
  et~al.}{2011}]{Mizuta_etal_2011ApJ...732...26M}
{Mizuta} A.,  {Nagataki} S.,   {Aoi} J.,  2011, \mn@doi [\apj]
  {10.1088/0004-637X/732/1/26}, \href
  {http://adsabs.harvard.edu/abs/2011ApJ...732...26M} {732, 26}

\bibitem[\protect\citeauthoryear{{Morsony}, {Lazzati}  \& {Begelman}}{{Morsony}
  et~al.}{2007}]{Morsony_etal_2007ApJ...665..569M}
{Morsony} B.~J.,  {Lazzati} D.,   {Begelman} M.~C.,  2007, \mn@doi [\apj]
  {10.1086/519483}, \href {http://adsabs.harvard.edu/abs/2007ApJ...665..569M}
  {665, 569}

\bibitem[\protect\citeauthoryear{{Morsony}, {Lazzati}  \& {Begelman}}{{Morsony}
  et~al.}{2010}]{Morsony_etal_2010ApJ...723..267M}
{Morsony} B.~J.,  {Lazzati} D.,   {Begelman} M.~C.,  2010, \mn@doi [\apj]
  {10.1088/0004-637X/723/1/267}, \href
  {http://adsabs.harvard.edu/abs/2010ApJ...723..267M} {723, 267}

\bibitem[\protect\citeauthoryear{{Nagakura}, {Ito}, {Kiuchi}  \&
  {Yamada}}{{Nagakura} et~al.}{2011}]{Nagakura_etal_2011ApJ...731...80N}
{Nagakura} H.,  {Ito} H.,  {Kiuchi} K.,   {Yamada} S.,  2011, \mn@doi [\apj]
  {10.1088/0004-637X/731/2/80}, \href
  {http://adsabs.harvard.edu/abs/2011ApJ...731...80N} {731, 80}

\bibitem[\protect\citeauthoryear{{Nagakura}, {Suwa}  \& {Ioka}}{{Nagakura}
  et~al.}{2012}]{Nagakura_etal_2012ApJ...754...85N}
{Nagakura} H.,  {Suwa} Y.,   {Ioka} K.,  2012, \mn@doi [\apj]
  {10.1088/0004-637X/754/2/85}, \href
  {http://adsabs.harvard.edu/abs/2012ApJ...754...85N} {754, 85}

\bibitem[\protect\citeauthoryear{{Nakar} \& {Sari}}{{Nakar} \&
  {Sari}}{2012}]{Nakar_Sari_2012ApJ...747...88}
{Nakar} E.,  {Sari} R.,  2012, \mn@doi [\apj] {10.1088/0004-637X/747/2/88},
  \href {http://adsabs.harvard.edu/abs/2012ApJ...747...88N} {747, 88}

\bibitem[\protect\citeauthoryear{{Pe'er}}{{Pe'er}}{2008}]{Peer_2008ApJ...682..463P}
{Pe'er} A.,  2008, \mn@doi [\apj] {10.1086/588136}, \href
  {http://adsabs.harvard.edu/abs/2008ApJ...682..463P} {682, 463}

\bibitem[\protect\citeauthoryear{{Rees} \& {Meszaros}}{{Rees} \&
  {Meszaros}}{1994}]{RM_1994ApJ}
{Rees} M.~J.,  {Meszaros} P.,  1994, \mn@doi [\apjl] {10.1086/187446}, \href
  {http://adsabs.harvard.edu/abs/1994ApJ...430L..93R} {430, L93}

\bibitem[\protect\citeauthoryear{{Rees} \& {M{\'e}sz{\'a}ros}}{{Rees} \&
  {M{\'e}sz{\'a}ros}}{2005}]{RM_2005ApJ...628..847R}
{Rees} M.~J.,  {M{\'e}sz{\'a}ros} P.,  2005, \mn@doi [\apj] {10.1086/430818},
  \href {http://adsabs.harvard.edu/abs/2005ApJ...628..847R} {628, 847}

\bibitem[\protect\citeauthoryear{{Sironi} \& {Spitkovsky}}{{Sironi} \&
  {Spitkovsky}}{2011}]{SS_2011ApJ...726...75}
{Sironi} L.,  {Spitkovsky} A.,  2011, \mn@doi [\apj]
  {10.1088/0004-637X/726/2/75}, \href
  {http://adsabs.harvard.edu/abs/2011ApJ...726...75S} {726, 75}

\bibitem[\protect\citeauthoryear{{Soderberg} et~al.,}{{Soderberg}
  et~al.}{2008}]{Soderberg_etal_2008Natur_short}
{Soderberg} A.~M.,  et~al., 2008, \mn@doi [\nat] {10.1038/nature06997}, \href
  {http://adsabs.harvard.edu/abs/2008Natur.453..469S} {453, 469}

\bibitem[\protect\citeauthoryear{{Thompson}}{{Thompson}}{1994}]{Thompson_1994MNRAS}
{Thompson} C.,  1994, \mnras, \href
  {http://adsabs.harvard.edu/abs/1994MNRAS.270..480T} {270, 480}

\bibitem[\protect\citeauthoryear{{Th{\"o}ne} et~al.,}{{Th{\"o}ne}
  et~al.}{2011}]{Thoene_etal_2011Natur_short}
{Th{\"o}ne} C.~C.,  et~al., 2011, \mn@doi [\nat] {10.1038/nature10611}, \href
  {http://adsabs.harvard.edu/abs/2011Natur.480...72T} {480, 72~(T11)}

\bibitem[\protect\citeauthoryear{{Usov}}{{Usov}}{1992}]{Usov_1992Natur}
{Usov} V.~V.,  1992, \mn@doi [\nat] {10.1038/357472a0}, \href
  {http://adsabs.harvard.edu/abs/1992Natur.357..472U} {357, 472}

\bibitem[\protect\citeauthoryear{{Zhang} \& {Yan}}{{Zhang} \&
  {Yan}}{2011}]{Zhang_Yan_2011ApJ...726...90Z}
{Zhang} B.,  {Yan} H.,  2011, \mn@doi [\apj] {10.1088/0004-637X/726/2/90},
  \href {http://adsabs.harvard.edu/abs/2011ApJ...726...90Z} {726, 90}

\bibitem[\protect\citeauthoryear{{Zhang}, {Woosley}  \& {MacFadyen}}{{Zhang}
  et~al.}{2003}]{Zhang_2003ApJ...586..356Z}
{Zhang} W.,  {Woosley} S.~E.,   {MacFadyen} A.~I.,  2003, \mn@doi [\apj]
  {10.1086/367609}, \href {http://adsabs.harvard.edu/abs/2003ApJ...586..356Z}
  {586, 356}

\bibitem[\protect\citeauthoryear{{Zhang}, {Woosley}  \& {Heger}}{{Zhang}
  et~al.}{2004}]{Zhang_2004ApJ...608..365Z}
{Zhang} W.,  {Woosley} S.~E.,   {Heger} A.,  2004, \mn@doi [\apj]
  {10.1086/386300}, \href {http://adsabs.harvard.edu/abs/2004ApJ...608..365Z}
  {608, 365}

\bibitem[\protect\citeauthoryear{{Zhang} et~al.,}{{Zhang}
  et~al.}{2011}]{Zhang_etal_2011ApJ...730..141Z}
{Zhang} B.-B.,  et~al., 2011, \mn@doi [\apj] {10.1088/0004-637X/730/2/141},
  \href {http://adsabs.harvard.edu/abs/2011ApJ...730..141Z} {730, 141}

\makeatother
\end{thebibliography}

\label{lastpage}

\end{document}